%% file: main.tex
\documentclass[10pt]{iopart} 

\RequirePackage{graphicx}
\usepackage{tabularx}
\newcolumntype{Y}{>{\centering\arraybackslash}X}
\usepackage[dvipsnames]{xcolor}
\RequirePackage[colorlinks = false]{hyperref}
\hypersetup{linkbordercolor = NavyBlue,citebordercolor = CornflowerBlue}
\usepackage{cite}
\usepackage{float}
\usepackage{enumitem}
\usepackage[normalem]{ulem}
\usepackage[acronym]{glossaries}

\newcommand{\ket}[1]{\left|#1\right\rangle}

\newcommand{\be}{\begin{equation}}
\newcommand{\ee}{\end{equation}}
\newcommand{\bea}{\begin{eqnarray}}
\newcommand{\eea}{\end{eqnarray}}

\makeatletter
\newcommand{\thickhline}{%
    \noalign {\ifnum 0=`}\fi \hrule height 1pt
    \futurelet \reserved@a \@xhline
}
\makeatother

\input{glossary}

\begin{document}

\title[A Framework for Quantum Curriculum Transformation]{A Framework for Curriculum Transformation in Quantum Information Science and Technology Education}

\author{Simon Goorney$^{1,2,\dagger}$, Jonas Bley$^{3,\dagger}$, Stefan Heusler$^4$, Jacob Sherson$^{1,2,*}$}

\address{$^1$ Department of Management, School of Business and Social Science, Aarhus University}
\address{$^2$ Niels Bohr Institute, University of Copenhagen, Denmark}
\address{$^3$ Department of physics and Research Center OPTIMAS, RPTU Kaiserslautern-Landau, 67663 Kaiserslautern, Germany}
\address{$^4$ Institute for Didactics of physics, Münster university, Wilhelm-Klemm-Str. 10, D-48149 Münster, Germany}
\address{$^\dagger$ These authors contributed equally to this work and share first authorship}
\address{$^{*}$ Correspondence: sherson@mgmt.au.dk}

\noindent{\it History\/}: Received: 21st August 2023, Revised: 6th December 2023


\begin{abstract}
     The field of \acrfull{qist} is booming. Due to this, many new educational courses and university programs are needed in order to prepare a workforce for the developing industry. Owing to its specialist nature, teaching approaches in this field can easily become disconnected from the substantial degree of science education research which aims to support the best approaches to teaching in \acrfull{stem} fields. In order to connect these two communities with a pragmatic and repeatable methodology, we have synthesised this educational research into a decision-tree based theoretical model for the transformation of \acrshort{qist} curricula, intended to provide a didactical perspective for practitioners. The \acrfull{qctf} consists of four steps: 1. choose a topic, 2. choose one or more targeted skills, 3. choose a learning goal and 4. choose a teaching approach that achieves this goal. We show how this can be done using an example curriculum and more specifically quantum teleportation as a basic concept of quantum communication within this curriculum. By approaching curriculum creation and transformation in this way, educational goals and outcomes are more clearly defined which is in the interest of the individual and the industry alike. The framework is intended to structure the narrative of \acrshort{qist} teaching, and with future testing and refinement it will form a basis for further research in the didactics of \acrshort{qist}.
\end{abstract}


\section{Introduction}\label{sec:intro}

The field of \acrfull{qist}, although originating in physics, is gaining increasing interdisciplinary popularity \cite{Asfaw_2022}. This is presently inducing a significant rise in the number of degree programs and courses in this field \cite{Aiello2021-we,Kaur2022-vf}, which requires a didactical perspective which is currently significantly lacking, due to its rapid development and specialist nature. As a result, it is of great importance that the results of didactical research are incorporated into quantum curriculum creation and transformation at the present time. There are also several immense challenges that the field presents: Its interdisciplinarity and complexity makes content preparation difficult as many concepts require perspectives from different fields. Additionally, many current teaching methods are suited to learners of particular academic backgrounds, stemming from disciplines such as physics and computer science. However, learners come to the \acrshort{qist} classroom from many fields which means that these approaches must be adapted.\cite{PhysRevPhysEducRes.18.010150}.

Given the importance of widespread, progressive education and training in \acrshort{qist}, the European Quantum Technology Education community (QTEdu) have developed practical guidelines for educators in non-formal contexts, and in the high school environment \cite{Goorney2022-yv,Andreotti2022-rn}. However, the most urgent need for quality education is in the university context, where graduates of Bachelor and Master programs in \acrshort{stem} fields will make up the emerging quantum workforce \cite{Kaur2022-vf}. More advanced topics in \acrshort{qist} are particularly challenging to teach, due to the specialist knowledge required \cite{10.1117/1.OE.61.8.081803}. As a result, the majority of teaching faculty are based in physics, Information Sciences, or Engineering, without access to significant pedagogical support \cite{PhysRevPhysEducRes.18.010150}.

\begin{figure}[htb]
    \centering
    \includegraphics[width=\columnwidth]{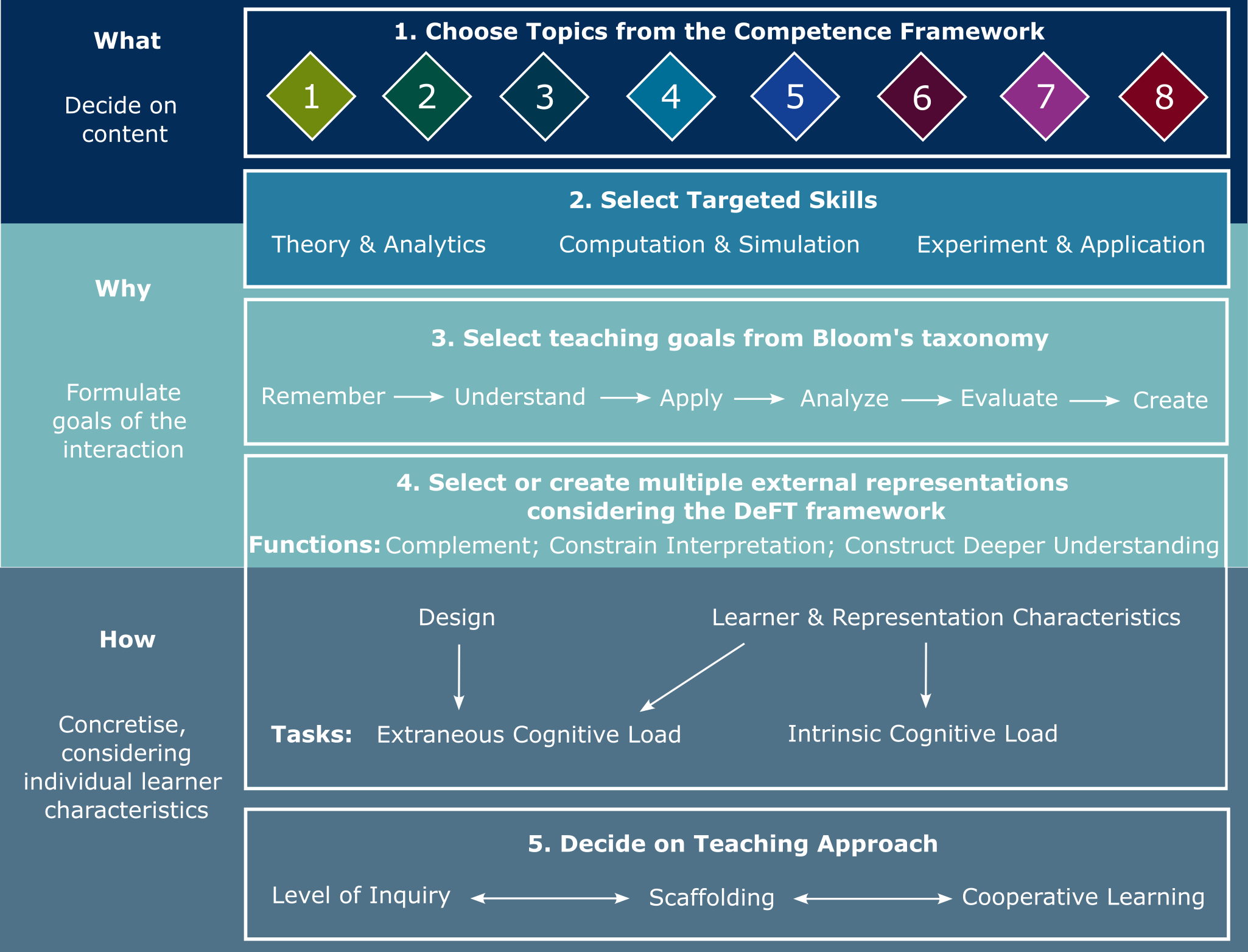}
    \caption{The Quantum Curriculum Transformation Framework (QCTF), a decision-tree based methodology for designing and transforming educational material and methods in \acrshort{qist}. It is based on a What - Why - How - approach, starting with the European Competence Framework for Quantum Technologies \cite{Greinert2023-ck} and concretising towards specific teaching approaches that are based on didactical research.}
    \label{fig:QCTF}
\end{figure}

Many of these practitioners are not grounded in knowledge of those core educational principles which may be background for high school teachers. For this reason, and in order to promote dialogue between didactics and educators, there needs to exist advice that is easy to follow. Additionally, this should be industry-needs oriented as preparing the quantum workforce for work in the industry is a major intent of many quantum courses \cite{9733176,Gerke_2022,Greinert2023-za}. For this purpose, in this work we present a didactical teaching framework, the \acrfull{qctf}, based on a decision tree which can be used to guide practitioners through teaching design, from topic selection to the specific teaching approach based on existing didactical theory. It should be emphasised that subsections of the framework may also be used in order to support specific aspects of pedagogical preparation, even if the complete framework is not applicable in every case. In addition, it may be extended and modified when presented with new research results. The framework is summarised in Fig. \ref{fig:QCTF}.

In the level of school education, there exists a storied tradition of research on the use of pedagogical content mapping, \cite{Angeli2013-oo,Baxter1999-hh}, as a means of aligning learning material with teaching methods which are most optimal for promoting different kinds of learning outcomes. However, this approach is rarely taken at the university level, and never in the field of \acrshort{qist}. There are many differences between higher education and secondary school in the complexity and the interdisciplinarity of content, in the time scopes and time available to study specific subjects and, perhaps most importantly, in the learners. Although at the university level, \acrshort{qist} has attracted attention toward research into student understanding \cite{Meyer_2023}, and development of novel teaching methods \cite{PhysRevPhysEducRes.18.010122}, there is little research available into how best to build up curricula from these components. This article, and the \acrshort{qctf} we propose, aims to address this. 

The paper is structured as follows: In Sect. \ref{sec:cfw}, we summarise the basis for choosing educational content (the ``what"), which makes use of the European \acrfull{cf} for Quantum Technologies \cite{Greinert2023-ck,greinert2024european}. Then, three targeted skills areas are introduced in Sect. \ref{sec:skills}, as both a motivation and a means of communicating the content chosen to teach. In Sect. \ref{sec:blooms}, a modified Bloom’s taxonomy is described for use in defining learning goals. In Sect. \ref{sec:deft}, Ainsworth's \acrfull{deft} framework is summarised and applied to the \acrshort{qist} context in Sect. \ref{sec:deft_app}. Lastly, we show a possible explicit teaching approach to quantum teleportation based on our framework in Sect. \ref{sec:example}. Throughout this work, we will stick to examples to ground the \acrshort{qctf} and convey the meaning of the didactical theory in the \acrshort{qist} context. Quantum teleportation is chosen as a topic specifically because it makes use of various concepts central to \acrshort{qist}: entanglement, measurements and unitary operations in three-qubit systems and has many applications \cite{Hu2023-dz}. In order to make the ordering of the framework's sections more explicit and intuitive for practitioners, we roughly structure it into a ``what" - ``why" - ``how" format, in which we first choose content to be taught, then formulate specific teaching goals for the interaction, and then consider the specific teaching approaches which best address these goals while also respecting the diversity of student backgrounds. As such, our approach is akin to a \gls{backwards} process \cite{Wiggins2005-oa}.

\section{Competence Framework}\label{sec:cfw}

\begin{figure}[htb]
    \centering
    \includegraphics[width=\columnwidth]{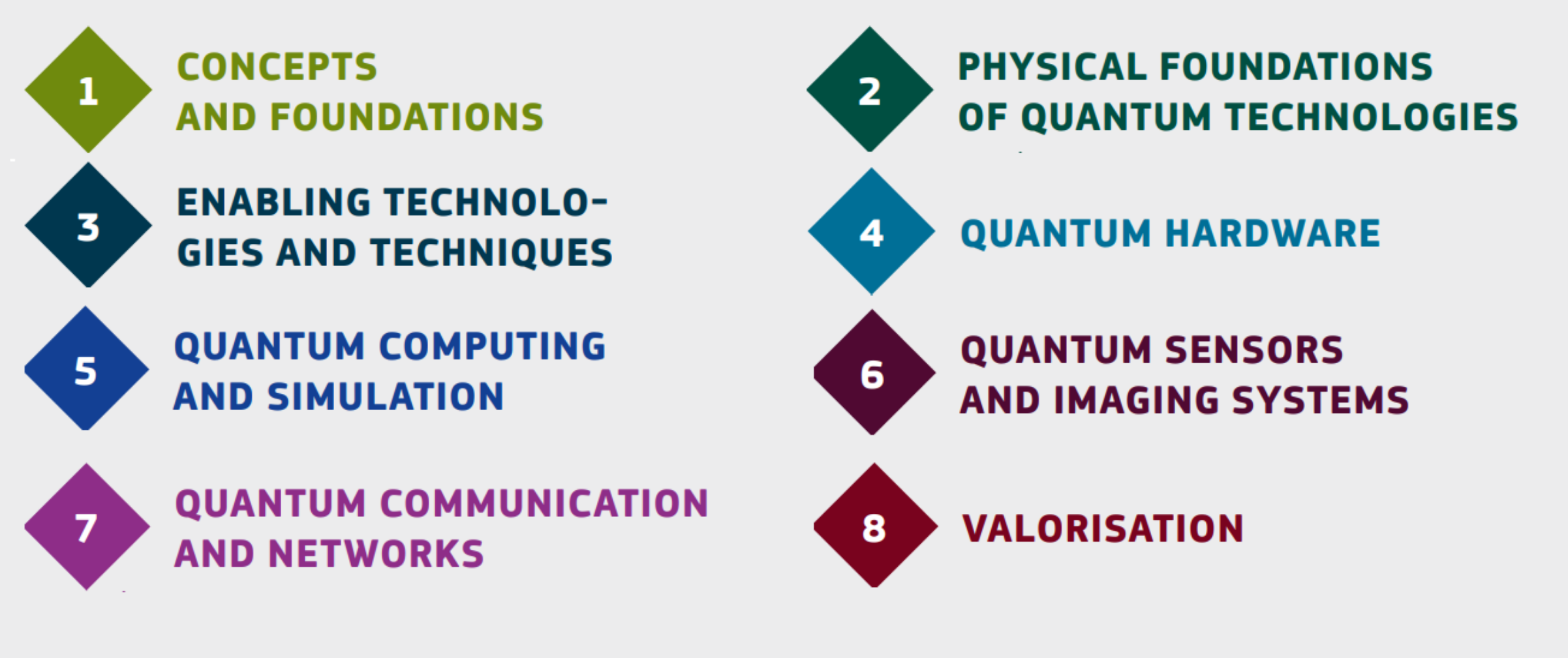}
    \caption{The 8 domains of the European Competence Framework for Quantum Technologies, covering Quantum Background (domains 1 and 2), Core Device Technologies (domains 3 and 4), and \acrshort{qt} systems and applications (domains 5-8). Adapted from \cite{Greinert2023-ck}.}
    \label{fig:cf_domains}
\end{figure}

The European \acrfull{cf} for Quantum Technologies is a tool designed by the Quantum Flagship, the EU funding program for \acrshort{qist}, intended as a common language for communicating skills in the field. The framework has utilised a Delphi method to identify key content according to industry needs \cite{Greinert2023-za}, and is presently available in version 2.0 \cite{Greinert2023-rz}, providing a structured overview of the field. It may be used to describe skills required for jobs, learning outcomes from courses, and content of entire degree programs. The \acrshort{qctf} builds on the \acrshort{cf} as a backbone, providing a means to determine content selection when designing educational initiatives. Each topical domain is assigned a number 1-8, covering Quantum Background (domains 1 and 2), Core Device Technologies (domains 3 and 4), and \acrshort{qt} systems and applications (domains 5-8). Within each domain, topics are assigned numbers such that a complete map can be generated of knowledge in \acrshort{qist}. An overview of the \acrshort{cf} domains is shown in Fig. \ref{fig:cf_domains}. 

\begin{figure}[htb]
    \centering
    \includegraphics[width=\columnwidth]{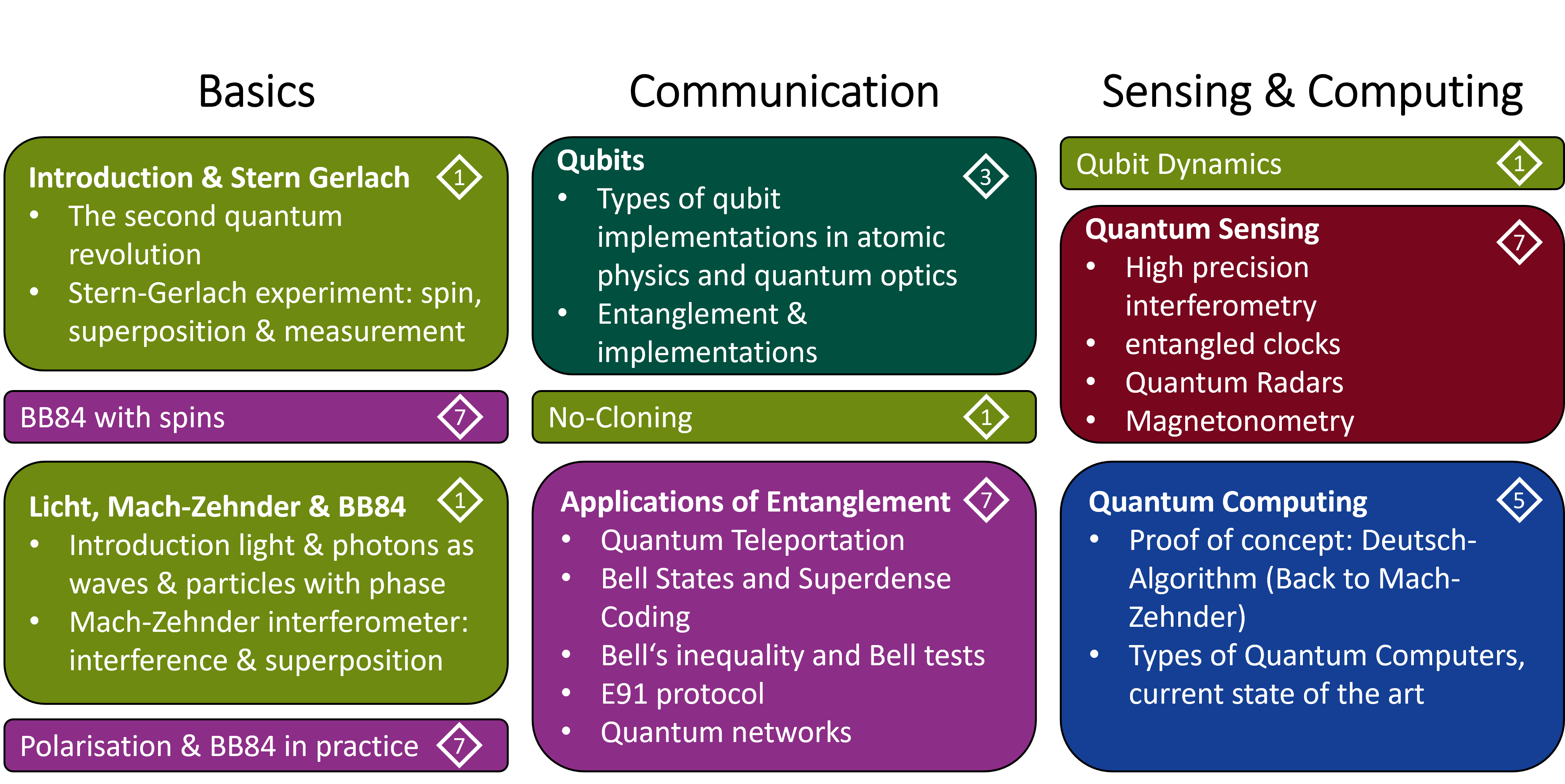}
    \caption{Application oriented example curriculum coloured using the Competence Framework \cite{Greinert2023-ck}. These three modules could be part of, e.g. a one- or two-week workshop covering the basics of \acrshort{qist}, quantum communication and giving an outlook into quantum sensing and computing.}
    \label{fig:curriculum}
\end{figure}

For use in designing curricula, we present an application-oriented example of a small curriculum with topics chosen from the \acrshort{cf}, focusing on the basics of quantum technologies and quantum communication, in Fig. \ref{fig:curriculum}. 

That plain knowledge transfer, e.g. in the form of purely a traditional lecture is not effective is well known \cite{Deslauriers2011-cv,Freeman2014-bj}. Therefore, planning courses should not stop after it has been decided which concepts to teach but rather should go further, considering first the skills that the learners should achieve in the context of the chosen concepts. We classify these skills in the following.

\section{Targeted Skills}\label{sec:skills}

In the context of higher education on quantum technology, we see possible targeted skills in three main areas: Theory \& Analytics, Computation \& Simulation and Experiment \& Real World. We emphasise that these targeted skills are those engaged in different approaches to teaching topics in all eight domains of the \acrshort{cf}. We do not present these as a classification scheme for concepts, which is in fact the role of the \acrshort{cf} itself \cite{Greinert2023-za}. Rather it is possible for many concepts to engage several or all of the “targeted skills”. Quantum Teleportation, for example, is a rich topic which can be approached with a view to developing each of these three targeted skills, while other topics may be limited to developing a subset of these skills. 

\acrshort{qist}, in many cases, allows or even necessitates an interdisciplinary view. For example, quantum computing algorithms are significantly restricted by the hardware that they run on. Gate times and fidelities limit circuit depth and general complexity of possible algorithms, which can be summarised by performance benchmarks \cite{Lubinski2023-qu}. This leads to the practical problems of finding the right quantum computer for a specific algorithm \cite{Salm2020-ba} and, in addition, determining which algorithms work well on a given hardware architecture \cite{Bluvstein2022-mf,Zhang2022-as}. This means that expertise in both areas is required to make justified decisions. As is also pointed out in \cite{Greinert2023-za}, generally one would address industry needs by having skills in all three of these skill areas.

\subsection{Theory \& Analytics}

When targeting theoretical skills, one targets the skill of derivation and proof of theoretical ideas within \acrshort{qist} like the no-cloning-theorem or the no-communication theorem. Instead of just being able to recite these theorems, learners ideally should understand the process of arriving at them. By doing so, they gain skills required to expand the given set of ideas, the theoretical framework of \acrshort{qist}, built up on complex euclidean spaces or so-called Hilbert spaces \cite{Watrous2018-ck,Bengtsson2006-ep}, to generate new knowledge. In Sect. \ref{sec:blooms}, we concretise this using Bloom’s taxonomy.

Analytical skills refer to calculations involving numbers or variables. In \acrshort{qist}, e.g. the Dirac ket notation is popular to represent vectors and linear transformations in these complex vector spaces. As an example, the skill of calculating the effects of various quantum logic gates (linear transformations) within this notation or the equivalent vector notation can be targeted. We refer to, e.g. the textbooks \cite{Watrous2018-ck,Bengtsson2006-ep,Nielsen2011-in} for more details on \acrshort{qist} from this perspective.

\subsection{Computation \& Simulation}

When targeting computational skills, a focus lies on programming within \acrshort{qist}. This can be done in various different languages and libraries, e.g. in Python using a library like Qiskit \cite{qiskit} or Pennylane \cite{bergholm2022pennylane}, in C++, e.g. with the Intel Quantum Software Development Kit (QSDK) \cite{Guerreschi2020-fg} or in the standalone languages Open Quantum Assembly Language (OpenQASM) \cite{Cross2022-py}, Q\# \cite{Hooyberghs2022-mq} or Silq \cite{noauthor_2020-oa}, but there are also many more that are reviewed in, e.g. \cite{Heim2020-oj}. These languages can be deployed on real quantum hardware or simulators. Real quantum hardware tends to be noisy, especially for large systems of qubits and simulating them is exponentially difficult in the number of qubits \cite{PhysRevX.10.041038,Xu2023-mk}. Gate errors can also be simulated and analysed within these simulations \cite{PhysRevA.96.062302} as these are and will be much more relevant than in classical computing. This is where the field of quantum error correction starts \cite{Raussendorf2012-tv}. 

We refer to textbooks such as \cite{wong,Fundamentals} for more detail on \acrshort{qist} from this perspective.

\subsection{Experiment \& Real World}

Last but not least, \acrshort{qist} can be approached from an experimental perspective. Skills in this area can cover a vast array of planning and preparation of experimental setups, a deep understanding of the functioning of the devices that are used, practical skills like calibration and adjustment of these devices and data acquisition and analysis \cite{PhysRevPhysEducRes.19.010117}. In the context of quantum information science and technology, qubit platforms and types are of special interest for transport of quantum information (quantum communication), precise measurement of real world quantities like temperature or magnetic fields (quantum sensing) or universal and analogue quantum computing. Knowledge and skills required to get into one of these fields in these areas can be targeted with textbooks like \cite{technology,Ezratty2021-dj} and in practical lab courses. Typical challenges that are discussed in more detail in \cite{PhysRevPhysEducRes.19.010117} include evoking excitement and making quantum effects ``visible" with these challenges being tied to another. How to achieve the latter and further benefits, like gaining intuition and familiarity of quantum mechanics, gaining conceptual understanding and getting a sense for possible applications in the field of quantum technologies, are discussed in \cite{PhysRevPhysEducRes.19.020144}.

All of these three categories are vast in that they each define multiple fields of research. The sheer endless amount of content to choose from and skills that could be acquired is a challenge for educators in the field of \acrshort{qist} education. Due to this complexity, theoretical frameworks like the \acrshort{cf} are needed to aid in the selection process. With our framework, we aim to go beyond the level of pure content towards a practical guideline that is applied to curriculum transformation and creation by educators. This is shown further in the following sections.

\section{Learning Objectives \& Bloom's taxonomy}\label{sec:blooms}

The more specific question of which cognitive learning goals within these areas of expertise should be accomplished can be asked next. A classification of these learning goals can help for various reasons. First of all, it encourages aiming for learning goals beyond just remembering and understanding concepts. Secondly, the structure can help align the educator’s purposes with that of the learners, aligning the teaching process with valid performance tests \cite{Mcdaniel2010-oo}. In \cite{Cooper2014-bq}, the authors discuss this alignment for the purpose of designing empirical studies, arguing that if learning goals beyond memorisation should be achieved, ways to test this also have to be found. In \acrshort{qist} education specifically, there is a lack of standardisation of teaching material and measurement of learning outcomes \cite{PhysRevPhysEducRes.18.010150}.

\begin{figure}[htb]
    \centering
    \includegraphics[width=0.8\columnwidth]{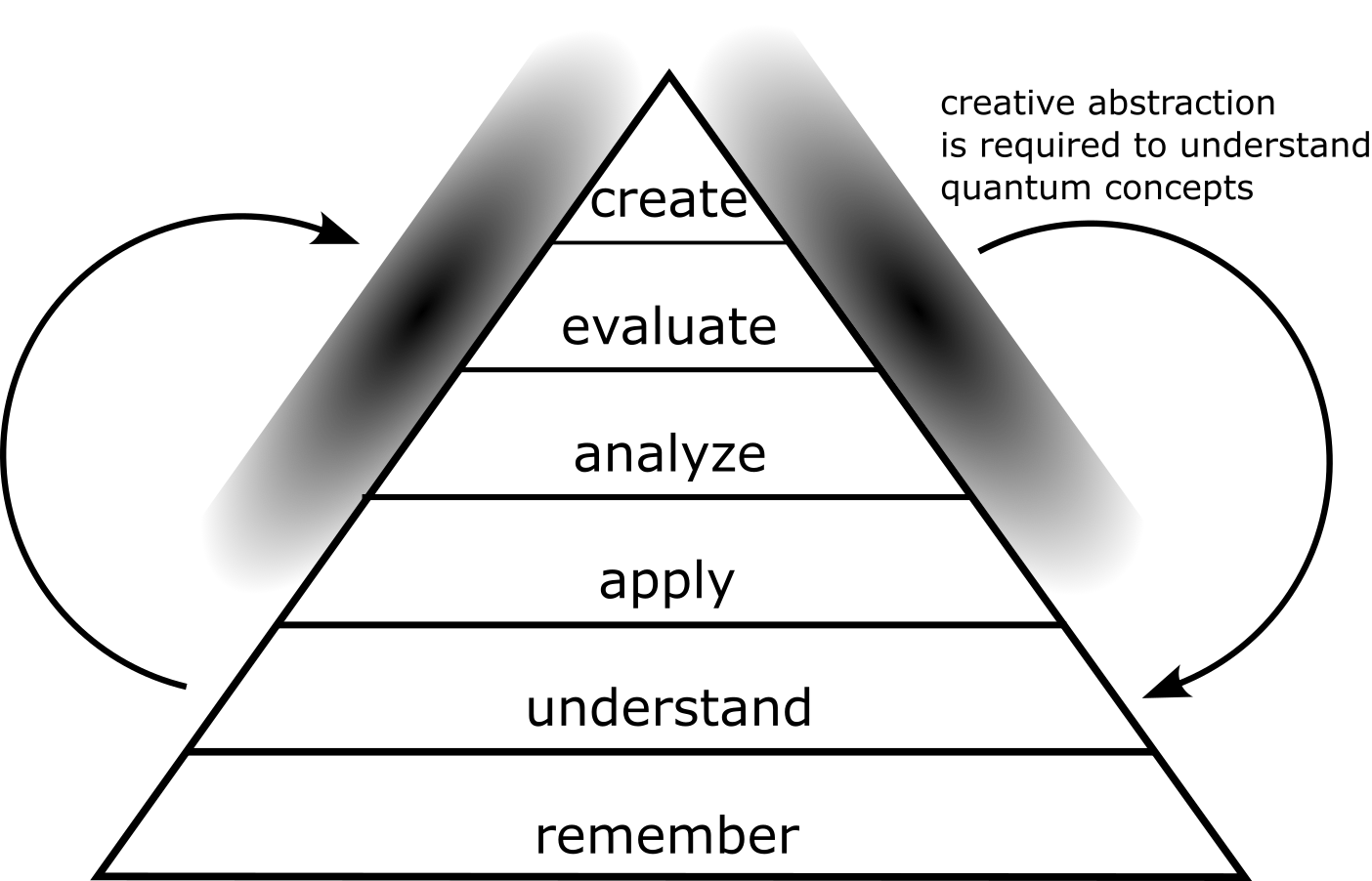}
    \caption{Revised Classical Bloom’s Taxonomy appropriate for use in describing learning outcomes in \acrshort{qist}, adapted from \cite{blooms_2005}. Since creative abstraction is required in order to understand basic quantum concepts, the relationship between Bloom’s thinking skills is non-hierarchical.}
    \label{fig:new_blooms}
\end{figure}

Bloom’s taxonomy is a classification scheme of such learning goals widely used in high school education. It comes in variations, but often six levels are defined based on increasing cognitive demand which represent lower order thinking skills to higher order thinking skills \cite{blooms_2005,Adams2015}. In the traditional taxonomy, higher order skills can only be reached if the previous skill is attained, defining a theoretical progression. This formulation of Bloom’s taxonomy is displayed in Fig. \ref{fig:new_blooms}.

Bloom’s taxonomy is used, e.g. in computer science education \cite{Masapanta-Carrion2018-gm} and is also explored in physics education \cite{Bhaw2020-de,EJPE}. Its wide use may be attributed to its simplicity and practicality, particularly in the high school setting where the hierarchical nature of the 6 levels can be used to set clear learning objectives for students with a variety of backgrounds and interests. Despite this, the taxonomy in its traditional form may come with some limitations, particularly when applied to the context of higher education. In \cite{Bouchee2022-oo}, Bhaw and Kriek point out that while cognitive demand associated with the six levels may indeed be hierarchical, learning situations in science education rarely involve only individual levels. Rather, they are inherently connected, for example when applied to problem-solving. In an evaluation problem for example, comparing two different approaches to address a problem requires understanding them both.

Similarly, in the context of \acrshort{qist}, we further re-interpret the taxonomy, for the following reason. The foundational concepts in \acrshort{qist} are well known to be abstract and disconnected from everyday life, far more so than many other areas of science \cite{Bouchee2022-oo}. As a result, the lower-order skills: understanding, and even basic knowledge, of concepts in \acrshort{qist} actually demand application of the higher-order skills. It is impossible to understand ideas such as superposition and entanglement, off which quantum technologies are based, without creative thought and imagination, as any possible realistic representation of these concepts are inherently non-everyday objects. This abstract nature also demands an inherent comparison to the classical world, and associated evaluative and analytical thinking. Our re-interpretation of Bloom’s taxonomy, as applicable in fields such as \acrshort{qist} in which the relationship between the cognitive skills involved is not purely hierarchical, is shown in Fig. \ref{fig:new_blooms}.

In the following, we provide example tasks in each of the three targeted skill areas for every level of the original Bloom’s taxonomy in the context of quantum teleportation. These don’t constitute an exhaustive list of such questions. Nevertheless, one can assume that the concept of quantum teleportation and its connection to \acrshort{qist} is thoroughly understood if all these tasks are accomplished by a learner.

\subsection{Bloom's taxonomy applied to Quantum Teleportation}

In the following, we show example tasks in the context of quantum teleportation for each of the three targeted skill areas (Sect. \ref{sec:skills} and for every level of Bloom's taxonomy) and cite references that can be consulted to approach these tasks. In this way, Bloom's taxonomy enables the creation of tasks for topics within \acrshort{qist} and structuring of the educator's intentions for student's learning outcomes. Bloom's taxonomy is often implemented using verb lists, where task-defining verbs are mapped to a corresponding level in Bloom's taxonomy, see, e.g. \cite{Shabatura_2014,Newton2020}, with some overlap of verbs into multiple levels, which is discussed in \cite{doi:10.1177/00220574211002199}. In the following, we use selected verbs from these lists.

\paragraph{Theory \& Analytics}

\begin{description}
\item[Remember]\hfill \\
Name the quantum gates used in the quantum teleportation algorithm. \cite{Nielsen2011-in}
\item[Understand]\hfill \\ Describe the process of quantum teleportation abstractly (independent of the implementation) \cite{Nielsen2011-in}. Describe why the no-cloning-theorem is not violated \cite{Ball2017-nf}. Discuss why no faster than light communication is happening. \cite{Ball2017-nf}
\item[Apply]\hfill \\ Use the Bell basis to describe the quantum teleportation algorithm mathematically. \cite{Dur2017-qx}
\item[Analyse]\hfill \\ Examine the effect of decoherence on the resulting state in the quantum teleportation algorithm. \cite{liu2022efficient,Bang_2018}

Examine the effect of quantum teleportation if it is applied to a qubit that is itself entangled in a multi-qubit system. \cite{Halder_2007}
\item[Evaluate]\hfill \\ Argue whether Quantum Teleportation is an inherent advantage of quantum physics in respect to classical physics or if it is more of a workaround necessary due to no-cloning. \cite{Liu2020-bb}
\item[Create]\hfill \\ Design a quantum teleportation algorithm for states of two qubits. \cite{Rigolin2005-oc,Ritik} Design a quantum teleportation algorithm such that three parties can exchange quantum information with each other. \cite{SEIDA2021167784}

\end{description}

\paragraph{Computation \& Simulation}
\begin{description}

\item[Remember]\hfill \\ List the functions in the Qiskit library that are used in the quantum teleportation algorithm. \cite{Guerreschi2020-fg}
\item[Understand]\hfill \\ Add comments to explain each line of code in the quantum teleportation algorithm. \cite{noauthor_undated-zb}
\item[Apply]\hfill \\ Translate the Qiskit code to another Python library, like Pennylane. \cite{Silverman_2023}
\item[Analyse]\hfill \\ Compare the resulting fidelities of teleported states in various quantum computers for different qubit choices. \cite{Bang_2018,noauthor_undated-zb}
\item[Evaluate]\hfill \\ Select the best hardware to perform quantum teleportation on (being restricted, e.g. to the free IBM hardware). \cite{Guerreschi2020-fg,Johnstun2021-af}
\item[Create]\hfill \\ Program a Python notebook with the Qiskit library, implementing a quantum network using a starting node, two middle nodes and an end node, where quantum information is teleported from one qubit to the last using three auxiliary qubits and a stationary qubit at every node \cite{Huang2020}. Modify the teleportation algorithm using quantum error correcting code and/or substituting quantum gates for gate combinations with smaller errors. \cite{Thacker2015QuantumCE}
Improve error correction using quantum teleportation. \cite{anandu2019demonstration}
\end{description}

\paragraph{Experiment \& Real World}
\begin{description}

\item[Remember]\hfill \\ List the necessary optical components used in all-optical quantum teleportation. \cite{Bouwmeester1997-qo}
\item[Understand]\hfill \\ Describe the function of the optical components within the quantum teleportation algorithm. \cite{Bouwmeester1997-qo}
\item[Apply]\hfill \\ Sketch an experimental design of quantum teleportation incorporated into a quantum cryptography protocol to extend the range of this protocol. \cite{Yin2020-nc}
\item[Analyse]\hfill \\ Differentiate Quantum Teleportation from other quantum optical algorithms shown in, e.g. \cite{Zubairy:01}. What are the differences and similarities in experimental setups?
\item[Evaluate]\hfill \\ Summarize the advantages and challenges of different areas of application (urban areas, ground-satellite, …) to weigh these areas against each other in the context of quantum communication. \cite{Ren2017-sp,Valivarthi2016-qh,Liu2020-lp,Ren2017-fu,Valivarthi2016-mq,Joshi2020-ym}
\item[Create]\hfill \\ Design an experiment to teleport photon states to  atomic particles. \cite{Sherson2006} Given an algorithm for teleportation of two-qubit states, extend the single qubit experimental design to teleport two-qubit states. \cite{Zhang2006-dc}
\end{description}

\noindent Having decided on tasks that should be covered according to Bloom's taxonomy, now is the time to answer \textit{how} exactly this is to be done, a question that can be answered by the educator and given as requirement to the students or by the students themselves. For example, the task ``Describe the process of quantum teleportation abstractly" can be answered descriptively (using words) symbolically (using mathematics), using graphical illustrations or any combination of these. Alternatively, students could be asked to ``translate" between these representations, for example to explain a schematic of the teleportation protocol. This is discussed in the following section.

In addition, some of these tasks may seem to difficult to answer straight away, especially some of the tasks in the \textit{Create} category. In these cases, proper support should be given to students (a practice commonly referred to as \gls{scaffolding} \cite{Belland2017} that we also briefly discuss in Sect. \ref{sec:scaffolding}). With this support and given enough time, it is possible that students find satisfying solutions for these tasks.

\section{Representations}\label{sec:deft}

An important part of the question of why and how to teach is why and how to represent the chosen concepts. By representation, here we mean anything external (as opposed to internal or mental), visual or auditory that is used to convey a concept \cite{Salimpour2021-cd}. Most of the time, multiple external representations (\acrshort{mer}s) will be used in science education as, for example, text will be combined with equations or figures. In the \acrshort{qist} context, \acrshort{mer}s are used in order to convey scientific ideas and concepts, like the Dirac bra-ket notation, vector and matrix notations, circuit diagrams or diagrams of experimental set ups. Understanding and working with these representations is part of the scientific workflow and necessary to be able to communicate concepts within the \acrshort{qist} community and can therefore be seen as an essential learning goal of \acrshort{qist} curricula.

\begin{figure}[htb]
    \centering
    \includegraphics[width=\columnwidth]{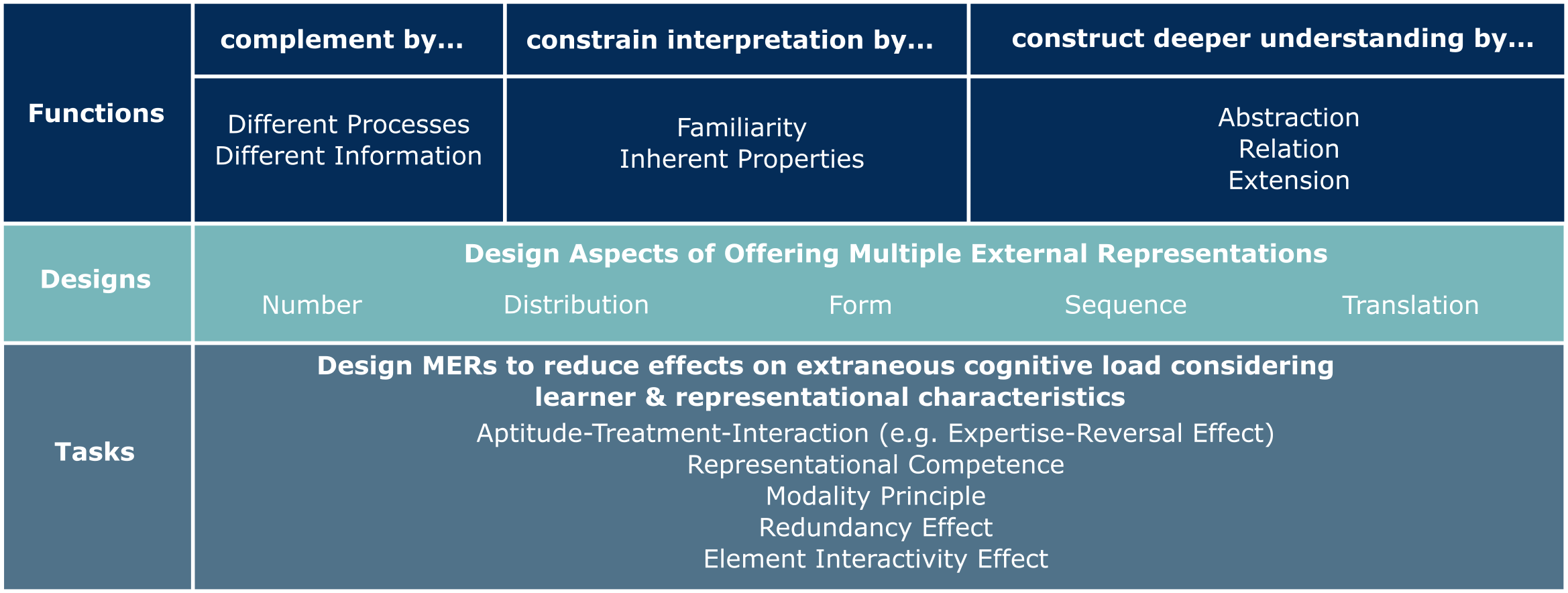}
    \caption{The \acrshort{deft} framework \cite{AINSWORTH2006183} adapted for the purpose of this work for the use of \acrshort{mer}s. Firstly, the functions describe various reasons for \textit{why} to use \acrshort{mer}s. Then, the \acrshort{mer}s should be designed (utilising the five design aspects) such that \acrshort{ecl} is reduced, considering learner characteristics and the listed effects \cite{Plass2010-zt}. How this is best done in practice in various \acrshort{qist} contexts is mostly an open research question.}
    \label{fig:deft}
\end{figure}

The \acrfull{deft} Framework \cite{Ainsworth1999-nn} is a theoretical framework for the use of \acrshort{mer}s that will be summarised here for our purpose. It consists of three layers to consider when combining representations: Designs, functions and tasks. While designs and tasks consider the how of representing information, the function layer constitutes the why. Therefore, in this work, we will firstly cover the function layer in order to line up the \acrshort{deft} framework with the taxonomy presented here. The DeFT framework is visualized in \ref{fig:deft}.

\subsection{Functions}
The \textit{Function} layer within this framework describes three main roles that \acrshort{mer}s can play: Firstly, to \textit{complement}, which means to provide various perspectives and information on the same concept so that learners can benefit from the specific advantages of each representation and to support different processes, i.e. strategies and individual differences of learners. For example, the Bloch sphere provides learners with an additional strategy of reasoning about the effect of quantum gates geometrically, as rotations on the sphere \cite{9951299}, complementing calculating in the computational basis typical Dirac notation. Another possible function \acrshort{mer}s is to \textit{constrain interpretation} i.e. to specify ambiguous or complex information, e.g. coming from text, for example with symbols or mathematical formulas. This can be an inherent property of the representation, or can be by familiarity of the learner with the representation that is used. And lastly to \textit{construct deeper understanding}, i.e. to support learners in achieving new insights on a concept which would be difficult with only one representation \cite{Ainsworth1999-nn}. The Bloch sphere would be another example of this as it visualises two complex numbers, utilising the redundancy of the global phase, to enable new visual insights on single-qubit quantum systems, sometimes even leading to scientific advances, e.g. in the context of Rabi oscillations \cite{PRXQuantum.2.040354,Karli2022}. In these various ways, different \acrshort{mer}s should be chosen carefully and given as tools to support learners in achieving the levels in Bloom’s taxonomy that the educators want them to achieve. For example, in order to select the best hardware for a given quantum algorithm, one could analyse coupling maps with gate errors like the one in \cite[Fig. 5]{couplinmaps}, while, on the other hand, this representation is not suitable to describe the mathematics of any quantum algorithm.

\subsection{Designs}
The \textit{Design} layer constitutes the \textit{how} of the use of \acrshort{mer}s. It describes the number of representations used, the way the information is distributed, the form of the \acrshort{mer}s (i.e. pictures, text, animations, sound, equations or graphs), the sequence (i.e. the order in which the representations are shown), and the translation (i.e. indications of relations between the representations, for example visual cues using colouring or spatial proximity). \acrshort{mer}s should be designed to reduce unnecessary cognitive tasks as described in the following.

\subsection{Tasks}
Last but not least, the \textit{Task} layer describes the cognitive tasks that learners need to overcome in order to benefit from the \acrshort{mer}s used. Cognitive tasks depend on characteristics of the representations (e.g. addressed sensory channel, modality, level of abstraction, type or dimensionality \cite{AINSWORTH2006183}) and on characteristics of the learners. For example, learners should gain representational competence, i.e. they need to understand the various Design aspects of the depicted \acrshort{mer}s, in order to benefit from them \cite{Scheid2018-jh}. Cognitive tasks are analysed within the \acrfull{clt} \cite{Kirschner2002-qj,Plass2010-zt}. The core idea is that ``Learning, reflected by performance change, requires working-memory capacity" \cite{Kirschner2002-qj}. Psychology research has uncovered that the human brain works on a limited amount of such temporary working memory, a short term storage, that is used to store information acutely necessary to perform a task \cite{BADDELEY197447,richardson1996working}. The working memory is often seen as consisting of two channels: a channel for visual and spatial information and a channel for audio information \cite{LOGIE1991105,Lehnert2006}. Further, it was shown that it is beneficial for learning to make use of both these memory channels by giving visual and audio information at the same time, for example in a typical presentation, as opposed to giving information in written words and pictures, i.e. using the same working memory channel for different information \cite{Mayer2003-pj}. This is referred to as the \gls{modality}. 

Another known effect of \acrshort{clt} is the redundancy effect suggesting that redundant material can hinder learning rather than having positive effects \cite{Kalyuga2014-ei}. The redundancy effect, e.g. suggests that experts, i.e. learners with high previous knowledge, could be hindered by the use of \acrshort{mer}s when the different representations don’t provide them with additional information or processes. This is referred to as the \gls{expertise} \cite{doi:10.1207/S15326985EP3801} as one form of the more general \gls{aptitude} \cite{https://doi.org/10.1002/tea.3660210804}. The \gls{splitattention} \cite{chandler1992} suggests that \acrshort{mer}s should be integrated to one single source of information rather than split either spatially or temporally. Another point to consider when using \acrshort{mer}s is that learners need to translate between these representations. This translation will be harder the more the representations differ in format and operators like various levels of abstraction or supported strategies \cite{Ainsworth1999-nn}. In \acrshort{qist}, a typical example is that of the use of vector notation and Dirac notation in course content design. If learners aren't good at translating between these representations, even if they are somewhat familiar with both, the use of both at the same time may lead to increased cognitive load and may therefore worsen learning outcomes \cite{AINSWORTH2006183}.

All of these effects have an influence on \Gls{extraneous} (\acrshort{ecl}) which is the cognitive load imposed on the learner by the way the information is provided, i.e. the \textit{Design}, and not the complexity of the information itself which is referred to as \Gls{intrinsic} (\acrshort{icl}) \cite{Mayer2005-ed}. When \acrshort{ecl} is reduced, learners can focus more on learning the content itself. Especially if the \acrshort{icl} of a given task is high, the negative effects of high \acrshort{ecl} can become stronger (\gls{element} \cite{Sweller2011_2}). Even though \acrshort{ecl} is usually seen as not desirable, some representations might be necessary to learn even though they are not the most efficient \acrshort{ecl}-wise, because they are widely accepted by the academic community and broadly used for communicating scientific concepts, an example of this being the Dirac notation. In the following, the \acrshort{deft} framework is applied to the \acrshort{qist} context in more detail.

\section{Applying the \acrshort{deft}-framework to \acrshort{qist} by the example of quantum teleportation}\label{sec:deft_app}
Quantum information science and technology is a field with a rich variety of forms of representations. To show this and how to apply the \acrshort{deft} framework to this space, we again refer to quantum teleportation and show examples of descriptive (i.e. with text), symbolic (i.e. mathematical) and graphical (i.e. with pictures and diagrams) visual representations of the process within the three different approaches. This is chosen as a practical classification of visual external representations, rather than a comprehensive one. As pointed out in, e.g. \cite{LaDue2015-cs}, creating a comprehensive classification of representations is a non-trivial endeavour that has been approached in, e.g. \cite{Black1962-ij}, a philosophical approach to models in language and maths, and \cite{Gilbert_2011}, where a general differentiation is made between verbal, mathematical and diagrammatic/pictorial models, similar to what we do in this paper. However, the work by Gilbert goes into much more detail. 

Models and representations are often defined using the other \cite{10.1093/oso/9780195087147.003.0009,2016_multimodal} and further exploring the semantic differences is beyond the scope of this paper. Rather than doing so or giving a comprehensive characterisation of such representations, this application of the \acrshort{deft} framework serves to show how possible representations can be categorised into the three targeted skill areas and point towards the thought processes that come with choice or creation of multiple suitable external representations using the \acrshort{deft} framework. As is laid out in \cite{Seufert_2004}, learners need to understand all the representations that they work with, all the given support on the syntactic and/or semantic level and how these representations connect to each other in order to gain deep understanding of the content. Some typical representations used in \acrshort{qist} are also discussed in \cite{10.1145/3481312.3481348,kushimo2023investigating} in the context of problem solving in Quantum Computing.

\subsection{Visual Representations of the Theory \& Analytics of Quantum Teleportation 
}
\paragraph{Descriptive} We provide a text/descriptive representation of the \textit{Theory \& Analytics} of quantum teleportation in \ref{sec:appA}. This is an abstract text describing the quantum teleportation protocol as a transfer of quantum information made possible due to the transitivity of entanglement. This description leaves some room for interpretation. For example, it describes a connection between entanglement and measurement that is not necessarily immediately visible. Therefore, it could benefit learners to constrain interpretation by providing symbolic and/or graphical representations to complement the description.

\paragraph{Symbolic} There are different ways to represent quantum teleportation symbolically. It is often represented in Dirac ket notation \cite{Nielsen2011-in}. Within this notation, one can use the computational basis $\ket{0}$ and $\ket{1}$. A more abstract formulation uses Bell states and Bell measurements, i.e. measurements in the Bell basis \cite{Vedral2002-bk}. Both are also shown in \ref{sec:appA}. There is a more general formulation with density matrices that can be suitable for a formal mathematical approach to the topic and can also be used for generalisations like the teleportation of mixed states and quantum gates \cite{Johnston2019-ov}. Within the \acrshort{deft} framework, the mathematical representation serves the function of \textit{complementing}, i.e. offering multiple perspectives on the same concept and providing learners with more strategies to explain these concepts to themselves or others by calculation. In addition, the mathematical language can \textit{constrain interpretation} by being less ambiguous than the text. Lastly, the formulation using Bell states and Bell measurements could \textit{construct deeper understanding} by abstraction which, e.g. could also be used in other contexts that benefit from a Bell state formulation like the investigation of Bell's inequality. On the other hand, learners could benefit from pre-existing familiarity with Bell states. Thus, it can be hypothesised that the text supported by mathematical representations helps learners understanding and problem solving in this context and that the abstraction in particular can help learners by enabling connections to other related contexts. Symbolic descriptions of quantum teleportation using the Dirac ket notation can be found in \ref{sec:appA}.

\begin{figure}[htb]
    \centering
    \includegraphics[width=0.49\columnwidth]{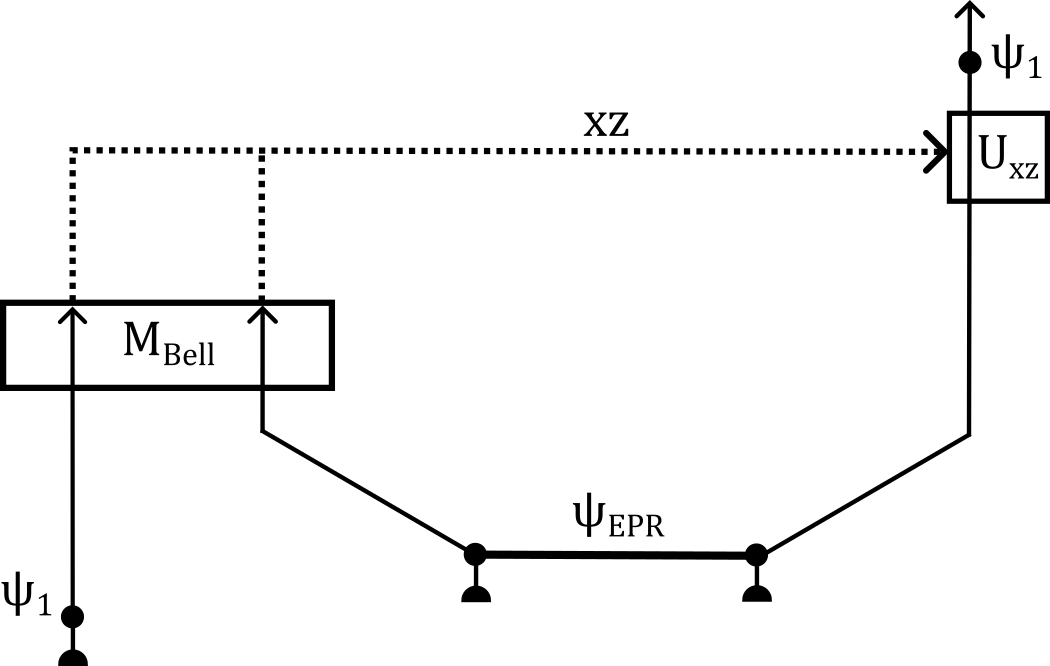}
    \includegraphics[width=0.49\columnwidth]{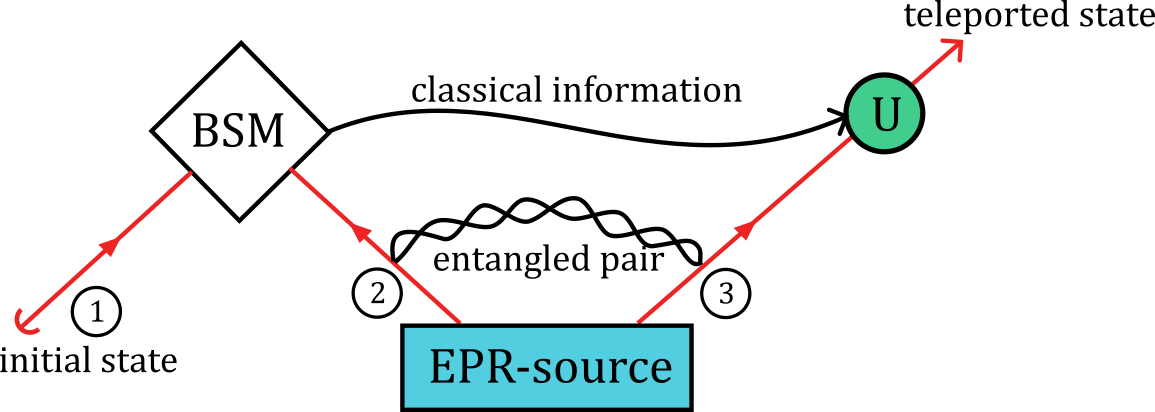}
    \caption{Schematic representations of Quantum Teleportation. These illustrations support a high level explanation of quantum teleportation and describe the flow of information, whether that information is quantum or classical as well as the role of the qubits in the protocol. To the left, the description leans more towards the mathematical perspective with states and classical information written as variables. The illustration to the right might be more suitable for learners that are more reluctant towards the mathematical symbols. Both combine multiple representations as both are arrow schematics supported by text or symbolic representations. The schematic to the left is adapted from \cite{Abramsky2004-ft} while the schematic to the right is adapted from \cite{Bouwmeester1997-qo}.
    }
    \label{fig:QT_schematic}
\end{figure}

\paragraph{Graphical} The theory of quantum teleportation can be visualised in abstract schematics (Fig. \ref{fig:QT_schematic}) \cite{coecke2004logic,Abramsky2004-ft}. Abstract representations like this allow for re-design and generalisation of processes like quantum teleportation and could provide benefits even for experts in that regard by supporting higher levels in Bloom’s taxonomy. A graphical representation providing an example for the creative abstraction needed to illustrate Bell-state measurements is provided by Fig. 4/5 in \cite{Dur2017-qx}. The typical ket notation can pose problems to students \cite{PhysRevPhysEducRes.18.010150}. To aid in understanding, the explicit analytics can, e.g. be graphically visualised using circle notation \cite{Johnston2019-ov} (see Fig. \ref{fig:QT_circle}) which complements the mathematical notation by providing the information of complex numbers in a graphical form. This is done with circles representing the basis states, inner circles of which the radius represents the absolute value of the corresponding coefficient and a line which represents the phase. Another possibility is introducing dimensionality, assigning every qubit an axis in space and attaching the amplitudes visually \cite{Just2023-cj}.

The SpinDrops program \cite{glaser_tesch_glaser} offers an alternative representation to the ones above, visualizing quantum states of up to three qubits in a different basis (akin to generalises Wigner functions \cite{PhysRevA.91.042122}) than the usual computational basis, putting a larger focus on quantum correlations, i.e. entanglement. An alternative way of visualizing entanglement properties in the computational basis is shown in \cite{Bley2023-na}, utilising the circle notation combined with dimensionality.

\subsection{Visual Representations of the Computation \& Simulation of Quantum Teleportation}

\paragraph{Descriptive} We provide computer code of Quantum Teleportation using the Python library Qiskit as a supplementary file. The code is annotated with description of what the functions do (complementing the Python code/providing support on the syntactic level \cite{LaDue2015-cs}), which step of quantum teleportation is being executed (providing support on the semantic level \cite{Seufert_2004}) and some other important notes like the impossibility to “read out” the quantum information of a qubit in the real world (providing deeper understanding).

\paragraph{Symbolic} As the gate-based language is often used in contexts describing the mathematical procedure, the same can be used for the purpose of supporting the computer code. Various functions and variables used within the code can be seen as the symbolic representation of the computation/simulation of quantum teleportation. This is a concise representation, albeit not easily understandable for someone not as familiar with the Python functions and the Qiskit library. Therefore, comments within the computer code are usually used to support the more concise formulation, improving readability. In addition, educators should make sure that learners have the necessary representational competence or know how to gain it using, e.g. the Qiskit documentation.

\paragraph{Graphical} A common visual representation of the code of quantum teleportation is the corresponding quantum circuit showing a line for every qubit and “boxes” for different gates. This picture can be annotated with the creation of the Bell state, Bell state measurement and final unitary operations (a question of design within the \acrshort{deft} framework). It is a more concise representation of the computer code and can therefore serve to constrain interpretation. In addition, the Bloch sphere can be used to describe the state $\ket{\psi_1}$ of the qubit to be teleported and compare it to the state  $\ket{\psi_3}$ of the qubit that the state is teleported onto in order to visualise whether the algorithm has worked. It can be used to more easily check whether the code works or not, giving access to the otherwise hidden information to provide learners with another strategy of bug fixing (in practice, this information is, ofcourse, actually hidden and can not be accessed by any means in a single measurement), complementing the Dirac ket notation output. On the Bloch sphere, it can be seen for example whether the X and Z gates are assigned to the wrong classical registers which could be more difficult when comparing the Dirac ket state output. An example of a Jupyter notebook that incorporates the quantum circuit representation and the Bloch sphere is shown in Fig. \ref{fig:QT_comp}. This could also be designed in the form of a computational essay, combining descriptive Educators need to make sure that learners have the necessary representational competence to use these tools.

\subsection{Visual Representations of Experimental Implementation \& Real World Application of Quantum Teleportation}

\paragraph{Descriptive} We provide a text representation of experiment and real world application of quantum teleportation in \ref{sec:appC}. It is described how Bell states between photons are created and how quantum teleportation can be verified in a proof-of-concept experiment by measuring correlations. This is a surface-level description with the aim of introducing quantum teleportation broadly while leaving open questions, for example how exactly spontaneous parametric down conversion \cite{PhysRevA.54.5349} or Hong-Ou-Mandel interference \cite{Walborn2003-ii} function. Explanations of these could be referenced, providing semantic support for the learners, but possibly increasing cognitive load significantly.

\paragraph{Symbolic} The underlying theory can be represented with polarisation letters $\ket{H}$ and $\ket{V}$ or arrows $\ket{\leftrightarrow}$ and $\ket{\updownarrow}$ as is done in, e.g. \cite{Bouwmeester1997-qo}. This means that the representation is designed such that the Dirac ket notation is related to the representation of horizontal and vertical polarisation, providing a connection between the two different representations and possibly benefiting learners in understanding of how the qubit is implemented optically.

\paragraph{Graphical} To describe real world quantum teleportation graphically, one can use a diagram including Alice and Bob to illustrate the process more abstractly, or show the more explicit optical experimental setup, both of which are commonly done, e.g. \cite{Nielsen2011-in,Bouchee2022-oo}. The more abstract representation can complement the information provided in a corresponding text on the overall process while the graphic of the optical experimental setup (see Fig. \ref{fig:QT_exp}) is more suitable to support a text describing the optical components and experimental setup and procedure.

\section{Teaching Approach}\label{sec:example}

Much of modern pedagogy is underpinned by Vygotsky’s constructivist theory of learning \cite{Vygotsky}. While in this article we will not dig into its development, which arose from the field of psychology, we will instead consider its implications for designing Quantum physics curricula. Put most simply, the chief implication of constructivism for the educator is that learners construct their own understanding, knowledge, and skills, through interaction with their environment. Crucially, this includes the social surroundings \cite{vonGlasersfeld1989} the core tenet of \textit{social constructivism}.

We argue that teaching must take into account this nature of learning, and thus propose three key considerations for educators which we call “teaching approaches''. These influence the environment in which learning is constructed. These are the chosen level of inquiry, and the degree of cooperative learning, and the degree of \gls{scaffolding} included. These are heavily intertwined as will become clear in this section.

Furthermore, all three pedagogical approaches are tunable factors which may be used by educators to adapt their teaching for \acrshort{qist} courses, which can be difficult to teach due to the diversity of student's disciplinary backgrounds. This includes different levels of prior competences in Mathematics and various MERs, programming, and many other areas. We describe the three approaches below. 

\subsection{Level of Inquiry}

\begin{figure}[htb]
    \centering
    \includegraphics[width=0.9\columnwidth]{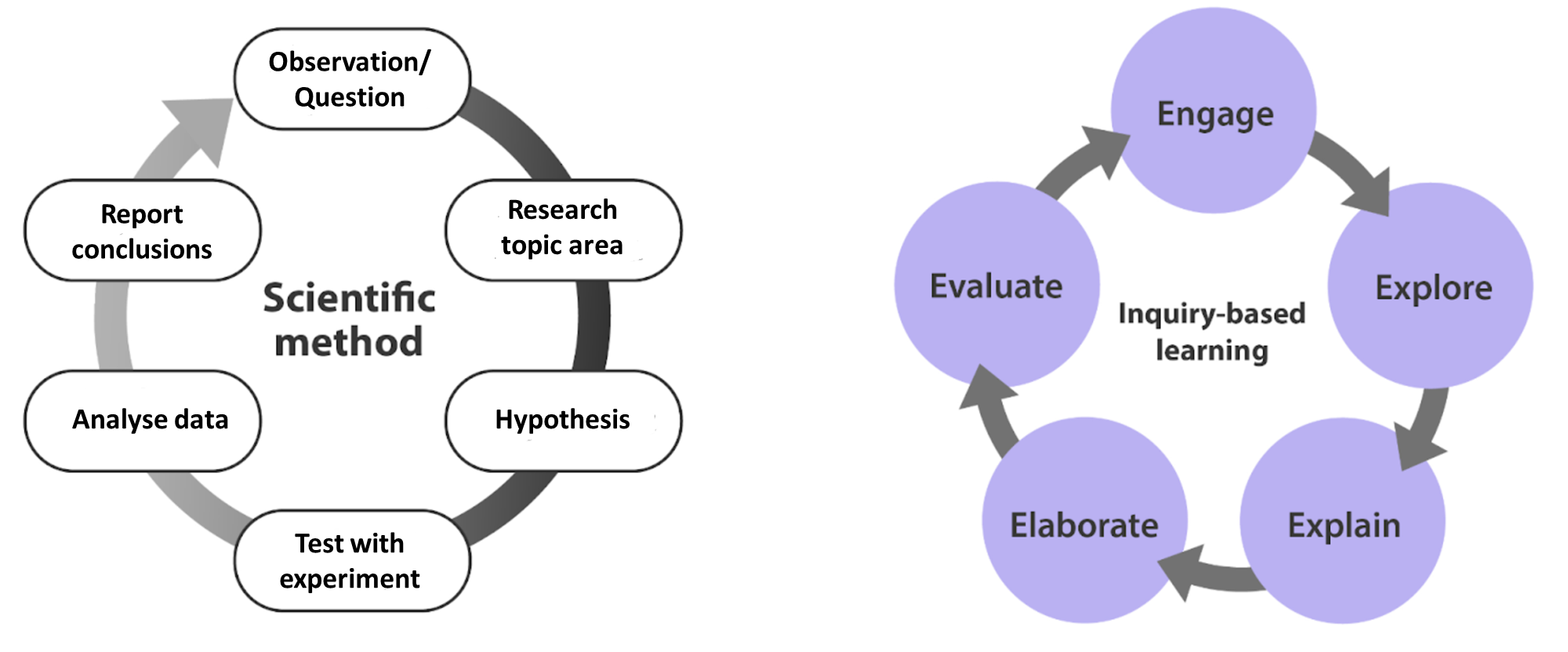}
    \caption{The scientific method (left) and the inquiry-based learning cycle (right). Both are cyclical processes, with a common structure (exploration, hypothesis-generation, testing/evaluation, and refinement.) As a result \acrshort{ibl} is a powerful means of teaching \acrshort{qist} concepts whilst also engaging scientific thinking skills.}
    \label{fig:cycle}
\end{figure}

Inquiry based education is a pedagogical approach in which learning is driven interactively \cite{Inquiry_2008}: by asking, and attempting to answer, questions. Inquiry is deeply embedded in science education because it is the nature of scientific study itself. There are many models of \Gls{inquiry} (\acrshort{ibl}) \cite{Bybee2006-oj,White1998-ls}, all sharing a cyclical and phasic nature. For example, the 5E’s cycle \cite{Bybee2006-oj} labels phases with Engagement, Exploration, Explanation, Elaboration, and Evaluation. These can be mapped onto the phases of the scientific inquiry cycle, as shown in Fig \ref{fig:cycle}. For this reason, \acrshort{ibl} may be not only beneficial, but in fact necessary for students to gain the skills needed for scientific thought. It should be noted that there exist other models of scientific inquiry, such as \textit{modelling-based inquiry} \cite{Gray_Campbell_2023} which describes the process by which a complex phenomena can be understood through constructing a scientific model. Regardless of the specific inquiry model referred to, the core process of attempting to uncover an answer to a question, in a cyclical manner, is the same. This gives rise to a class of inquiry based learning approaches. 

A core benefit of \acrshort{ibl} is that, as engaging the very process of inquiry is the goal of any interaction, it respects diversity of student backgrounds and prior learning to a greater extent than more teacher-led teaching methods. We thus advocate that in the interdisciplinary \acrshort{qist} environment, some degree of \acrshort{ibl} is essential. One may then ask what strategies can be taken to implement it in a teaching scenario. Pragmatism dictates a distinction between different ``levels of inquiry”, the most frequently used model of which is Bell and Banchi’s (Confirmation, Structured, Guided, Open) \cite{Banchi2008-lu}. 

Research suggests that conceptual understanding and subject knowledge are best fostered in more closed, structured inquiry based tasks \cite{Inquiry_2008}, whilst scientific thinking and inquiry skills are best fostered with more open, unstructured tasks \cite{White1998-ls}. This may be explained by the greater cognitive load associated with the more open levels of inquiry \cite{White1998-ls,Tanchuk2020-ay}, leaving less cognitive ability to develop conceptual understanding or commit ideas to long-term memory. For this reason, educators must choose carefully the level of inquiry in order to match their learning goals. In the field of \acrshort{qist}, there have been many recent examples of interactive tools which create possibilities for teaching approaches which span the four levels of inquiry \cite{Seskir2022-fn}. For example, the visual and gamified simulation tool for Quantum Computing, Quantum Odyssey \cite{Nita2021-dh}, is designed with a coherent structure of stand-alone scenarios addressing individual concepts, such as particular gate operations and communication protocols. These can be used to create more structured and confirmatory inquiry scenarios in which participants are both introduced to concepts and walked through them towards attaining particular targeted knowledge and understanding around the concepts. In addition, Odyssey has an editor mode in which users can create quantum algorithms from scratch using custom gates, therefore allowing for more open inquiry. The Quantum Composer tool is also effective for open inquiry, as it provides a blank slate interface in which staff and students may construct any one-dimensional quantum system and investigate its properties. In fact, the tool has been used to support research in real problems in \acrshort{qist} \cite{Sorensen2019-ci,PhysRevA.100.052314}, which demonstrates how this form of inquiry can support scientific thinking as a learning outcome. Both of these tools provide a means of visual representation of concepts in \acrshort{qist}, but it should be noted that \acrshort{ibl} models need not only use visualisation, but other forms of representation as described in Sect. \ref{sec:deft}. 

\subsection{Cooperative Learning}

Cooperative learning is an educational practice where learners work in small groups in order to achieve learning goals under the instruction of an educator. There is a large amount of empirical evidence that cooperation is more effective than competition and individualistic efforts in enabling higher achievement and productivity and that therefore, cooperative learning improves learning outcomes \cite{Johnson1981-qg,Johnson1991-ky,Johnson2000-tz}. This is only true, however, if cooperative learning is implemented correctly. In \cite{Johnson1991-ky}, five basic elements of effective cooperative learning are identified: (1) positive \gls{interdependence}, i.e. everyone is needed and contributes to the group's goal, (2) face-to-face promotive interaction, i.e. enabling constructive help and criticism within the group, (3) individual accountability and personal responsibility (e.g. to reduce diffusion of responsibility), (4) frequent use of interpersonal and small group social skills that should be promoted and rewarded and (5) frequent, regular group processing of current function, i.e. conjoint reflection on the group work. Findings also suggest that this depends on the level of inquiry, because more open ended questions are an effective tool to promote group discussion \cite{Cavinato2021-uj}. Therefore, the decision of how to incorporate cooperative learning into teaching is intertwined with the decision on the level of inquiry discussed previously.

There are findings that suggest that cooperative learning has advantages especially in an interdisciplinary context for tasks requiring input from multiple perspectives \cite{CORCHUELOMARTINEZAZUA2014INT}. Therefore, cooperative or team-working skills should be fostered as they are especially needed not only in general, but also more specifically in the interdisciplinary context of \acrshort{qist} \cite{Aiello2021-we,PhysRevPhysEducRes.16.020131,Venegas-Gomez2020-bn}. 
Findings suggest that diverse teams perform better than homogeneous teams under the right conditions \cite{Patricio2022-kv,asee_peer_30339}. One such condition is psychological safety, reducing dysfunctional social barriers within the group. However, the exact impact of cognitive diversity on team performance, i.e. the differences of the way people think which is supposedly influenced by academic background, is still an open research question \cite{Mello2015-zz,Liao2016-zw}. Part of promoting functional team work is improving student’s understanding that their individual previous knowledge and skills and also that of others are relevant to the group and the common goals that the group wants to achieve \cite{Nafukho2012-tm,Agrawal2012-qq} and therefore a ``diversity mindset” \cite{Van_Knippenberg2013-dd}. Diversity in academic background is one dimension of diversity such that one can argue that cooperative learning might be especially beneficial in the field of \acrshort{qist} with its diverse range of students' academic backgrounds \cite{PhysRevPhysEducRes.18.010150}.

In practice, more open tasks that require multiple perspectives are often tasks reaching into higher levels of Bloom’s taxonomy. These tasks might benefit more from cooperative learning methods than tasks on the lower levels of Bloom's taxonomy, because they allow cognitive diversity to be utilised more effectively. However, there is no clear evidence of this and it is very dependent on the specific context \cite{Huang2020}. Further, there exists evidence that teaching techniques like peer-to-peer instruction are beneficial also when trying to achieve lower levels of Bloom's taxonomy \cite{blooms_cooperative,Huang2020} and when applying Bloom's taxonomy in general \cite{10.1007/978-3-319-39672-9_1}. Nevertheless, making sure teams are diverse in academic background helps foster cooperative skills and improve group performance and learning when solving tasks with high \gls{interdependence} \cite{https://doi.org/10.1002/job.220}. Examining this effect is a possible direction of future empirical research in \acrshort{qist} education.

\subsection{Scaffolding} \label{sec:scaffolding}

\Gls{scaffolding} is a well known teaching practice that can be defined as interactive, temporary support given to learners for tasks that they otherwise would not be able to complete \cite{vandePol2010,Belland2014}. Here, the \textit{contingency}, i.e. interactivity or responsiveness, is what differentiates it from non-\gls{scaffolding} support \cite{vandePol2010}. \Gls{scaffolding} techniques include promoting student interest, controlling frustration, providing feedback, helping with task/subtask prioritisation, offering models of expert processes and questioning/encouragement of reflection \cite{vandePol2010,Belland2014}. \Gls{scaffolding} can be seen as essential to \acrshort{ibl} as a transfer of responsibility from educator to student is a key characteristic \cite{vandePol2010,doi:10.1080/00461520701263368}. Examples of how \gls{scaffolding} can be applied in \acrshort{ibl} in various \acrshort{stem} context include \cite{doi:10.1080/095006900412293,https://doi.org/10.1002/sce.10004,MAMUN2020103695,chemistry_scaffolding,Gross2021} and there exists extensive theory on how to do so \cite{noauthor_2005-hp,ecc169a7-bd4d-3c0c-8db5-abdb69812d52,Hsu2015}. \Gls{scaffolding} can be offered one-to-one, by peers; it can be computer-based or AI generated \cite{Belland2017,Umutlu2022}. In many different contexts, it was shown that students perform better when \gls{scaffolding} techniques were applied \cite{doi:10.1177/01655515030295004,doi:10.1080/00461520.2011.611369,CHEN20161201,10.1007/978-3-319-91464-0_10}, also when applying cooperative learning techniques and in the context of quantum physics \cite{onah2022effect}. In addition, \gls{scaffolding} techniques were shown to reduce anxiety and to have beneficial effects on motivation and reading comprehension \cite{AhmedAbdel2023}.

Especially in the \acrshort{qist} context, we argue that various \gls{scaffolding} techniques could be beneficial, because they consider individual academic backgrounds by definition. In addition, the concepts in \acrshort{qist} are quite abstract while the methods to deal with these concepts are often new to students, with possibly discouraging effect. Lastly, the field of \acrshort{qist} is new and therefore, knowledge acquisition often requires students to do their own research, reading scientific research articles that are not designed with them as a target audience in mind. \Gls{scaffolding} could, e.g. be provided in the form of providing on demand suitable visualisations, encouraging peer-to-peer instruction and cooperativity (as covered in the previous section), providing help with representations and how to perform calculations, concept maps (e.g. connecting content with content from different targeted skill areas or different domains of the \acrshort{cf}) aiding in research tasks by providing strategies and foundational material for understanding or close interactions with researchers (e.g. interviews).

\subsection{Teaching Approach to Quantum Teleportation}

In order to show how to put the framework to use, we present an example lesson on quantum teleportation using our framework. It is structured using the what - why - how approach and the corresponding steps in the framework.

\paragraph{What} This example teaching unit could be part of the example curriculum presented in Fig. \ref{fig:curriculum}. The goal of this curriculum is to introduce learners to the basics of quantum technologies, targeting primarily applications in the real world. Previous to the lesson, learners have been introduced to single-photon states and optical equipment like beam splitters and wave plates. They know how to use single photons to decrypt messages using the BB84 key distribution protocol. Then, they were introduced to the creation of entangled photon pairs via spontaneous parametric down conversion and a possible use of entanglement for efficient transport of classical information using superdense coding. During the latter, they learned the concepts of Bell states and Bell state measurements. Learners understand both the experimental implementation and description of these concepts using qubits and Dirac ket notation. During this example teaching unit, learners are introduced to quantum teleportation. To start the lesson, quantum teleportation can be introduced as a quantum protocol where -- in contrast to superdense coding that is designed to send classical information -- quantum information is transported over arbitrary distances.

\paragraph{Why} After this teaching unit, learners should understand the theory behind quantum teleportation and its experimental realisation in quantum optics. They are able to apply quantum teleportation in quantum networks in practice and can discuss applications, e.g. in quantum key distribution, comparing it to classical networks and classical key distribution in these networks. Learners can also evaluate the challenges that quantum networks pose and discuss possible ways of coping with these challenges. In terms of Bloom’s taxonomy, this means that the learners should reach the level of \textit{Understanding} in the \textit{Theory \& Analytics} area of expertise and reach the \textit{Evaluate} level in the \textit{Experiment \& Real world} area of expertise. This should enable them to understand practical problems within quantum communication and make justified decisions regarding various aspects of implementation of the quantum teleportation algorithm.

\paragraph{How} There are infinitely many possible ways to teach quantum teleportation with these prerequisites and goals in mind. However, the process of first deciding on learning goals and then thinking about how to achieve them, also called a \gls{backwards} process, is still much more constrained than an upwards design process, where you start with textbook contents or activities and figure out the learning goals from there \cite{Wiggins2005-oa}. We will discuss only one possibility that incorporates the ideas laid out in this paper. For this, we propose five steps in which the following inquiry is to be investigated: What is quantum teleportation and what role could it play in quantum networks? What are the advantages and challenges of such quantum networks in comparison to classical networks? The teaching steps, the corresponding targeted skill areas, levels of Bloom's taxonomy and activities are summarised in Tab. \ref{tab:teaching} and described in more detail in the following, including proposals for possible \gls{scaffolding} techniques.

\begin{table}[htb]\caption{\label{tab:teaching} The five steps of our example teaching approach to quantum teleportation summarised: In various skill areas, learners achieve the corresponding levels in Bloom's taxonomy (and the levels before) in the activities while supported by (optional) \gls{scaffolding} techniques.}
    \begin{tabularx}{\textwidth}{c|Y|Y|Y}
         Step & Skill Area & Up to this level in Bloom's taxonomy & Activity\\ \thickhline
         (i) & Theory & Understand &  Listening to short, conceptual presentation on quantum teleportation and discussion \\ \hline
         (ii)  & Analytics & Apply & Calculate necessary gates, choose proper representations\\ \hline
         (iii) & Experiment & Apply &  Plan an experiment and sketch an experimental setup\\ \hline
         (iv) & Real World Application & Evaluate & Discuss real world application of quantum teleportation in quantum networks in a presentation\\ \hline
         (v) & Computation \& Simulation or Experiment & Remember & Carry out experiments or outlook on further use cases of quantum teleportation in quantum computing \& programming the algorithm
    \end{tabularx}
\end{table}

\begin{enumerate}
    \item Learners are introduced to the quantum teleportation protocol with a description and a schematic like the one in Fig. \ref{fig:QT_schematic} via presentation. The general idea of a transfer of quantum information by transfer of entanglement is introduced. The presenter should keep principles and effects of cognitive load theory summarised in Sect. \ref{sec:deft} in mind. The learners are asked why the protocol does not utilise faster than light communication and why it does not contradict the no-cloning theorem in group discussion. To end the presentation, possible applications in quantum computing (entanglement swapping) and quantum communication (quantum repeaters) are touched on. The educator-student interaction can be scaffolded by the educator, including giving appropriate hints and direction and giving substantive feedback \cite{https://doi.org/10.1598/RRQ.37.1.4}.
    \item Learners are instructed to find out the explicit quantum gates that are needed for the quantum teleportation protocol (see \ref{sec:appA}). For this, they know that a Bell pair $\ket{\text{EPR}}$ has to be created, and $\ket{1}$ is part of a Bell state measurement by Alice with one of the two qubits in the Bell pair. The aim is to find the right unitary operations that Bob has to apply to the third qubit. For this, e.g. circle notation or dimensional circle notation could be used or just the Dirac ket notation. The choice of representation can be left to the learners (which representation is the most helpful for this purpose is an open didactical research question). Examples of possible graphical representations are shown in Fig. \ref{fig:QT_circle}. This task is probably not suitable for cooperative approaches, because it is very closed inquiry-wise. Here, further \gls{scaffolding} can be deployed by providing suitable representations and possibly giving advice on how to utilise them. 
    \item The third task is to plan an optical experiment to test whether quantum teleportation actually works in reality. For this, a schematic like in Fig. \ref{fig:QT_exp}, a list of necessary optical components and a descriptive text should be created, as well as a methodology and a prediction for the experiment. The learners are told to use the ideas of how to create an entangled photon pair and how to perform a Bell measurement. Therefore, they need to have or develop skills in the \textit{Apply} level of Bloom’s taxonomy regarding the previous topics of Bell state creation and measurements. They need to figure out how to make sure that it is clear that an entangled photon pair is created and how to prepare different initial states. The learners can try and predict Bob’s measurement outcome probabilities depending on Alice’s measurement outcomes. A problem in this experiment is that Bob has little time to adjust his experimental setup. To solve this, one could use ways of prolonging the path of Bob’s photon and finding ways to apply conditional optical gates as in, e.g. \cite{Valivarthi2016-mq} or, as in \cite{Bouwmeester1997-qo}, only consider the Bell state $\psi^-$ (coincidences of Alice’s two detectors) and discard all other cases. In that case, Bob’s measurement outcome is certain and he can check whether that is the case, referring to measurements at detector $p$. This task can be done in groups as its openness and the possibility to incorporate many technical details into the experiment can promote group discussion. Possible \gls{scaffolding} techniques include promoting positive \gls{interdependence} within these group discussions, providing a list of optical components that can be used with descriptions and providing feedback and hints in case of problems.
    \item Preparing and holding a presentation of the finished work as is usually done also in science and industry is a considerable final task. In this presentation that is again prepared and done in a group, learners discuss how to implement quantum teleportation into quantum networks, how these networks could look like and what these networks could be used for. They are tasked with discussing quantum key distribution over large distances where multiple network nodes are necessary. In addition, learners discuss the challenges of building such networks, especially decoherence, quantify these errors from the literature and explain what could be possible measures to cope with these challenges. They then evaluate the advantages and disadvantages in comparison to classical networks that are secured with classical key distribution. Groups should make a plan and split up, completing subtasks individually or in pairs and staying in contact. For example, one person could learn about classical networks and inform the others about how they function while another person learns about the details of spontaneous parametric down conversion, such that this information can be incorporated into the final presentation. This procedure makes sure that the group members are positively interdependent. Communication and social interaction should play a role in the working process and regular group meetings and discussions about the current status should be encouraged \cite{Johnson1991-ky}. Learners are also tasked to find or create suitable representations for conveying their thoughts and ideas. \Gls{scaffolding} techniques that can be used here are, again, promoting positive \gls{interdependence} by providing instruction on effective group work, propose possible structures and inquiries to deal with in the presentation (e.g. how would quantum networks function, what would they look like abstractly, or what challenges do they face), recommend literature and provide reading aid and encouraging the use of suitable representations.
    \item Outlook and possible next steps: If sufficient lab equipment is available, preparing this protocol and testing it in a real lab could be a possible next step. The amount of time this would take, however, can only be estimated by the educators. It could be a difficult and time consuming experiment to make work, depending on the available equipment and preparation done. Alternatively, the learners could be tasked with programming quantum teleportation in a programming language of their/the educators choice. An example of such a program is shown in Fig. \ref{fig:QT_comp}, using a Python notebook provided in the supplementary files. Different levels of fill-in code could be prepared for the students such that the different backgrounds in coding Python and Qiskit are considered (a form of \gls{scaffolding} \cite{Belland2014}). The code could be in the form of computational essays \cite{PhysRevPhysEducRes.15.020152,Odden_2020} which can be combined with \gls{scaffolding} \cite{BOdden_2021}. Then, learners would have reached at least the level of \textit{Understanding} in Bloom’s taxonomy from each of the three different perspectives that \acrshort{qist} offers.
\end{enumerate}

\section{Outlook and Discussion}\label{sec:outlook}

Given the substantial expansion in the number of courses and degree programs attempting to educate a workforce in the specialist field of \acrshort{qist}, it is crucial that careful attention is paid to how to do this most effectively. The development of the \acrshort{qctf} approach described in this article is intended as a guiding tool to be used for educators, in a pragmatic and digestible manner, to reduce an overwhelming library of educational research down to a single framework which may be used when designing educational material and classroom approaches.

The framework is structured in the manner of “what” - “why” - “how”, and thus breaks down curriculum construction into individual elements, which may be addressed with the support of didactical theory and tools. The first - which content to include - may be addressed using the Competence Framework, designed as a common language for content in \acrshort{qist}, and a major EU policy initiative which will become widespread in years to come. One must also select which skills are intended to be targeted in their education: be they theoretical, computational, or experimental. These skills are part of both the “what” and the “why”, as they are in themselves a learning outcome as skills to be developed through the education, but equally they are skills used akin to the knowledge used in addressing the areas of the \acrshort{cf} representing field-specific content.

Educators must then select the learning-outcome specific thinking skills they wish to address. Foundational to this approach is Bloom’s Taxonomy, which we argue in the field of \acrshort{qist} requires re-interpretation due to the strongly counter-intuitive nature of quantum concepts, but is still a useful guiding principle, albeit with less hierarchy in skill development. Given the content and targeted skills, educators designing material must next decide on the kind of representation they use to present concepts or explain to students such that they can represent concepts themselves. The \acrshort{deft} framework is the tool intended to support this, and we emphasise that the framework is a support structure which may be used to several degrees of depth. In the most pragmatic and simple application, one may choose tasks based on the representational competences they wish learners to acquire, for example using Dirac’s bra-ket notation when intending a mathematical approach, or using a simulation tool for a visualisation approach.

The final aspect of how the framework is applied is in choosing the teaching approach applied in the classroom. Here we consider the level of inquiry, ranging from more open and exploratory approaches, addressing more scientific thinking skills, to more guided ones appropriate for more directly targeting acquisition of field-specific knowledge. One must also consider how students interact with others in their environment, and cooperation in learning can be a key factor in determining the exact activities used for teaching, beyond traditional “chalk and talk” approaches which lack the benefit of a social underpinning. 

We emphasise that the framework is a guiding structure, intended to distil the results of educational research into a pragmatic approach for educators. Whilst most effectively applied in full, educators may also pick aspects from the framework to address individually - for example refining primarily their level of inquiry, or choosing topics using the \acrshort{cf}. The framework is also open to refinement and iteration based on community feedback, as more and more educational material is developed in this field, and in particular with an eye to a didactically-sound approach. One particular refinement which may be beneficial in the future is the inclusion of proficiency levels in the ``what" stage, while choosing content. Suggested proficiency levels were included in the Competence Framework version 2.0 \cite{Greinert2023-rz}, providing a rough equivalency between content difficulty and qualification level. Since then, the framework was refined on the basis of interviews with industry representatives, adding 9 qualification profiles that could further guide development of curricula \cite{greinert2024european}. The refinement of the proficiency levels, as the framework itself, we recommend for future work in the course of the flagship European Quantum Technology education projects, DigiQ \cite{DigiQ} and QTIndu \cite{QTIndu}.

\section*{Acknowledgements}

We thank Franziska Greinert from the TU Braunschweig for valuable discussion.

J.B. acknowledges support by the project QuanTUK at the RPTU in Kaiserslautern, supported by the Federal Ministry of Education and Research (FKZ13N15995).

\appendix

\section{Visual Representations of the Theory \& Analytics of Quantum Teleportation}\label{sec:appA}

Here, we categorise typical representations in \acrshort{qist} in three categories, but point out that this is neither a comprehensive categorisation of representations \cite{LaDue2015-cs} nor a comprehensive list of possible representations of quantum teleportation. These representations are suitable for introductory \acrshort{qist} courses targeting different skills in the three skill areas (Sect. \ref{sec:skills}).

\subsection{Descriptive} Quantum teleportation can be seen as a transfer of quantum information made possible by a transfer of entanglement \cite{Nielsen2011-in}. Qubit \#1 is in an arbitrary state that is to be teleported to qubit \#3. For this, qubit \#3 is entangled with qubit \#2 in a Bell state. Upon performing a Bell measurement on qubit \#1 and qubit \#2, one will get one of four possible results with equal probability. By doing so, quantum information of the qubit \#1 is transferred to qubit \#3. However, by definition of entanglement, the exact state of qubit \#3 after measurement depends on the measurement result. This way, quantum information is transported over arbitrary distances, but can only be retrieved after the classical information has been transmitted. This information is not copied, because the quantum information contained in qubit \#1 collapsed due to the measurement.

\subsection{Symbolic} The quantum teleportation algorithm can be seen as a gate-based computing algorithm and is explained similarly to here in many textbooks. The system starts in the state $\ket{\psi_i}$ that consists of a Bell state $\ket{\psi_{23}}$ and the state $\ket{\psi_1}$ of qubit \#1 to be teleported:

\begin{align}
    \ket{\psi_i} &=\ket{\psi_1}\otimes \ket{\psi_{23}} \\
    &= (\alpha\ket{0}+\beta\ket{1})\otimes (\frac{1}{\sqrt{2}}\ket{00}+\frac{1}{\sqrt{2}}\ket{11})\\
    &= \frac{\alpha}{\sqrt{2}}\ket{000}+\frac{\beta}{\sqrt{2}}\ket{001}+\frac{\alpha}{\sqrt{2}}\ket{011}+\frac{\beta}{\sqrt{2}}\ket{111}.
\end{align}

Alice entangles qubit \#1 and \#2 using a CNOT-gate and applies a Hadamard gate on qubit \#1 to prepare a Bell state measurement of qubit \#1 and \#2 to obtain the final state $\ket{\psi_f}$ before measuring:

\begin{align}
    \ket{\psi_f}&=H_1CNOT_{12}\ket{\psi_i}\\
    &=H_1 (\frac{\alpha}{\sqrt{2}}\ket{000}+\frac{\alpha}{\sqrt{2}}\ket{011}+\frac{\beta}{\sqrt{2}}\ket{110}+\frac{\beta}{\sqrt{2}}\ket{101}) \\
    &= \frac{1}{2} \left(\alpha(\ket{000}+\ket{100})+\alpha(\ket{011}+\ket{111})+\beta(\ket{010}-\ket{110})+\beta(\ket{001}-\ket{101})\right).
\end{align}

To show what happens upon measuring qubit \#1 and \#2, the state can be rewritten as

 \begin{align}
     \ket{\psi_f}=&\ket{00}\otimes (\alpha\ket{0}+\beta\ket{1}) \\
     +&\ket{01}\otimes (\beta\ket{0}+\alpha\ket{1})\\+&\ket{10}\otimes (\alpha\ket{0}-\beta\ket{1})\\+&\ket{11}\otimes (\beta\ket{0}-\alpha\ket{1}).
 \end{align}

Upon measuring, Bobs qubit \#3 will be in one of these four possible states and Bob needs to apply gates depending on the measurement result in order to obtain the initial state of qubit \#1:

\begin{align}
    00: \;&\alpha\ket{0}+\beta\ket{1} \\
    01: \;&\beta\ket{0}+\alpha\ket{1} \xrightarrow{X} \alpha\ket{0}+\beta\ket{1} \\
    10: \;&\alpha\ket{0}-\beta\ket{1} \xrightarrow{Z} \alpha\ket{0}+\beta\ket{1} \\
    11: \;&\beta\ket{0}-\alpha\ket{1} \xrightarrow{XZ} \alpha\ket{0}+\beta\ket{1}.
\end{align}

One can alternatively represent the initial state of quantum teleportation using the Bell basis:

\begin{align}
    \ket{\psi}_1\ket{\phi^+}_{23}= &\frac{1}{\sqrt{2}}(\alpha\ket{000}+\alpha\ket{011}+\beta \ket{100}+\beta \ket{111}) \\
    = &[\ket{\phi^+}_{12} (\alpha\ket{0}+\beta\ket{1})_3 \\
    = &\ket{\phi^-}_{12} (\alpha\ket{0}-\beta\ket{1})_3 \\
    = &\ket{\psi^+}_{12} (\beta\ket{0}+\alpha\ket{1})_3 \\
    = &\ket{\psi^-}_{12} (\beta\ket{0}-\beta\ket{1})_3]
\end{align}

Then, a Bell measurement is performed by Alice to achieve the desired result. This representation supports the more abstract explanation of quantum teleportation using the Bell basis \cite{coecke2004logic,Dur2017-qx} (see also Fig. \ref{fig:abstract_QT}). Another alternative is the use of the vector and matrix representation \cite{Nielsen2011-in}, but, considering these three-qubit vectors would have eight entries, this seems like an impractical solution. 

\subsection{Graphical}

\begin{figure}[htb]
    \centering
    \includegraphics[width=0.6\columnwidth]{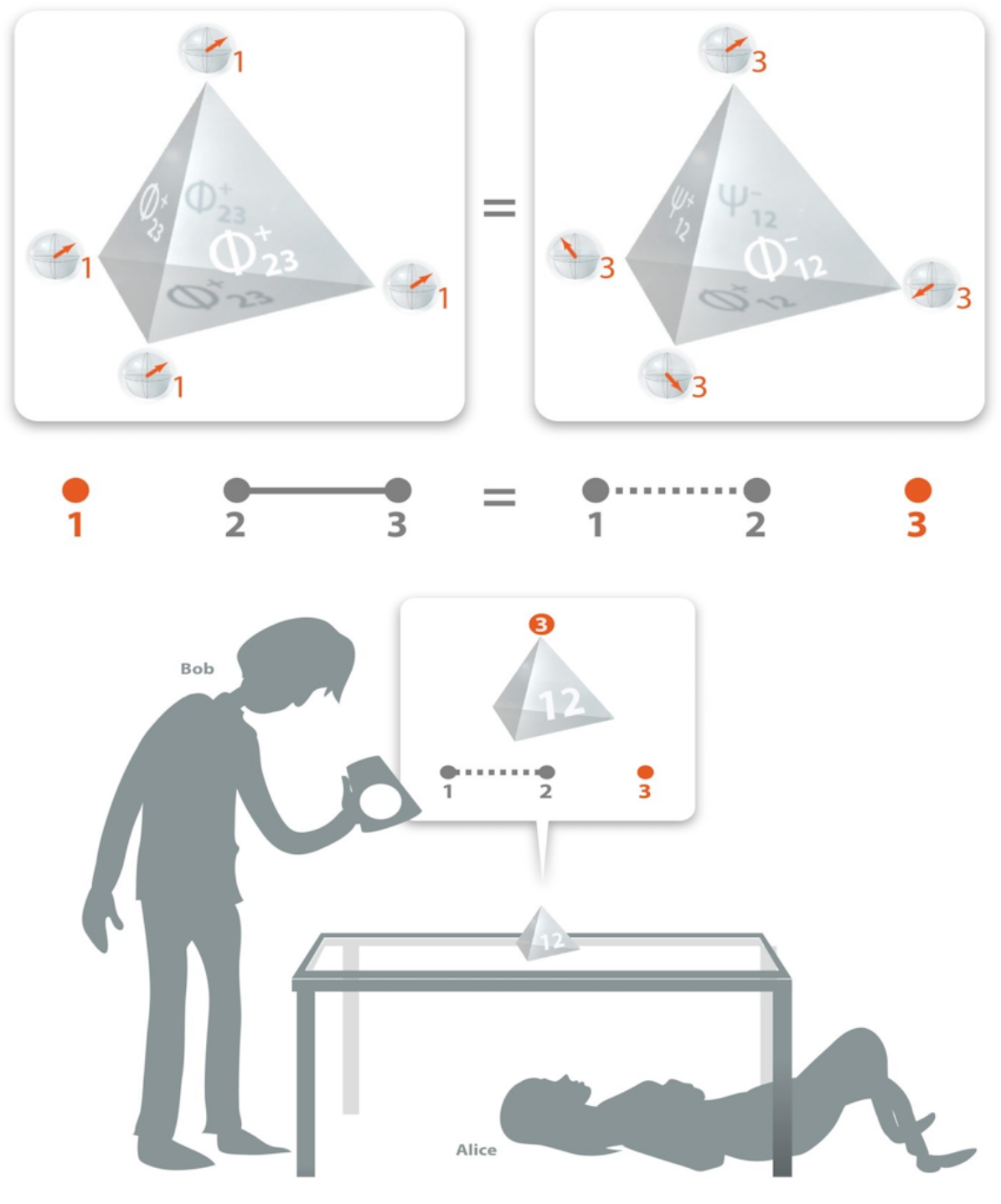}
    \caption{Creative abstraction of the mathematical relations relevant for quantum teleportation: There are two equivalent manners to represent the three-photon state $\ket{\psi_1}\ket{\phi^+_{23}}$ (see \ref{sec:appB}). In this visualisation of this abstract relation, Bob throws a pyramodial “dice” with four possible outcomes, corresponding to four possible Bell-states at Alice site, at the corresponding four possible teleported photon states at his side \cite{Dur2017-qx}. Once Alice informs Bob about her result, he knows which local unitary transformation he has to perform at his photon to teleport Alice original state $\ket{\psi_1}$  to his side. This approach is more abstract than the one in Fig. 8, focussing on the basis transformation from computational basis to the Bell basis that the explicit circle and dimensional circle notations don’t represent.}
    \label{fig:abstract_QT}
\end{figure}

\begin{figure}[htb]
    \centering
    \includegraphics[width=0.9\columnwidth]{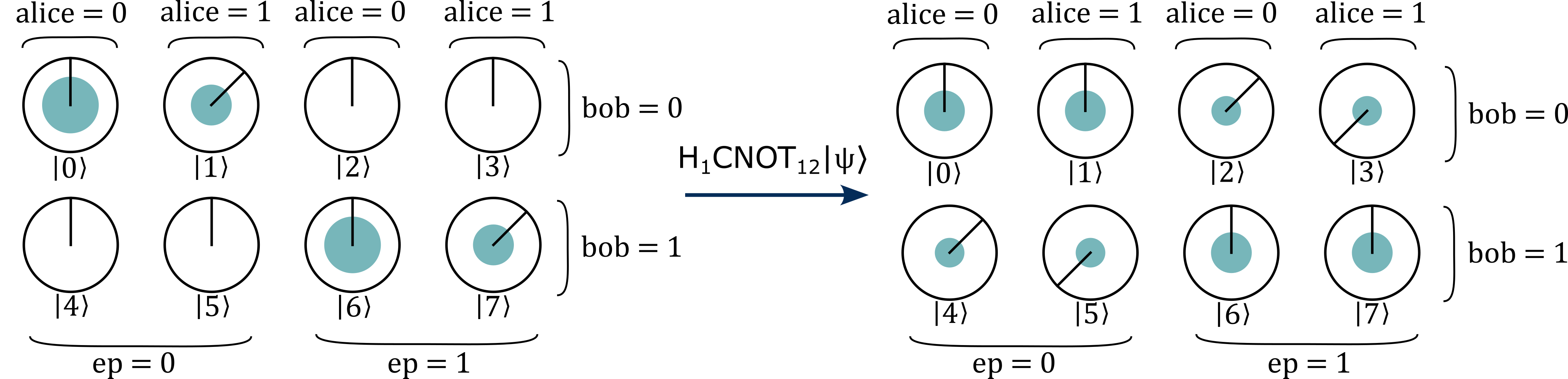}
    \includegraphics[width=0.85\columnwidth]{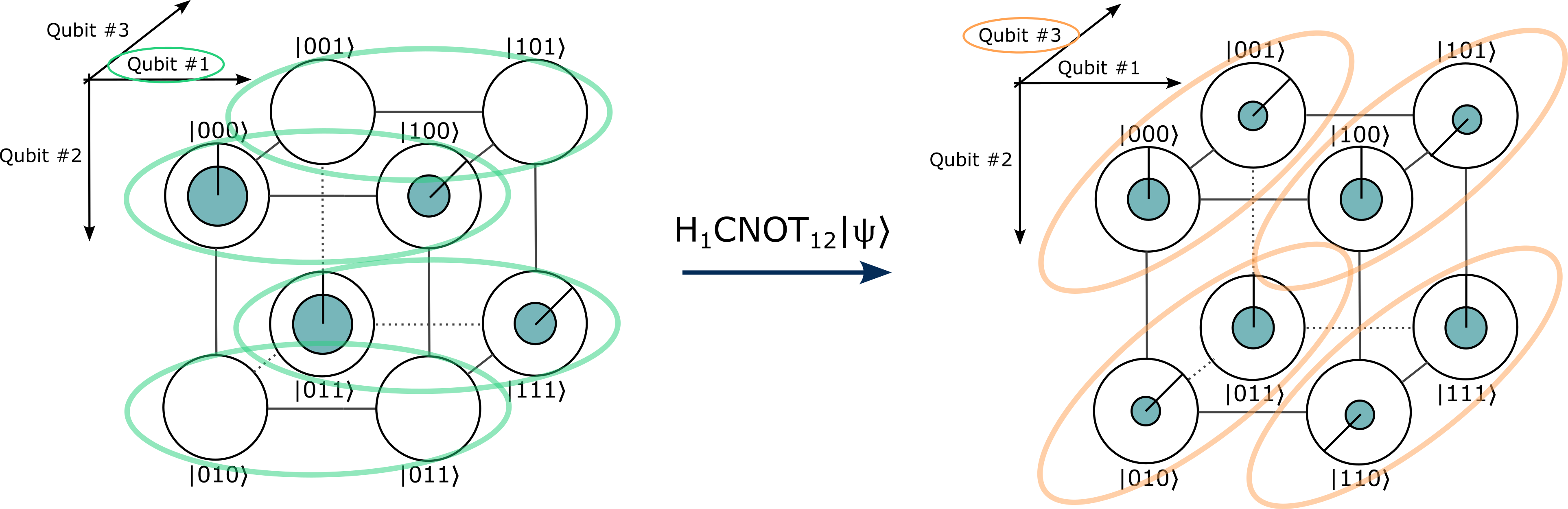}
    \caption{Graphical representations of an explicit mathematical example of the application of Alice's gates within the quantum teleportation algorithms (see \ref{sec:appB}), adapted from \cite{Johnston2019-ov,Just2023-cj,Bley2023-na}. This step is laid out more explicitly in all of these works. The illustrations could help learners gain understanding of the theory behind the protocol, provided they have the necessary representational competence. This can be done by visually laying focus on the information transfer between qubit \#1 and that of qubit \#3 or, in the case of the representation on the bottom, the information transfer from one axis to the other represented by the green and yellow ellipses \cite{Bley2023-na}.
    }
    \label{fig:QT_circle}
\end{figure}

Here we provide graphical representations of the application of Alice's gates in the quantum teleportation algorithm. In the more abstract Fig. \ref{fig:abstract_QT}, Alice's shift to the Bell basis ("looking" at the system differently) transforms the system such that Bob can receive the appropriate information, while the representations in Fig. \ref{fig:QT_circle} are more explicit and only represent the step in the computational basis.

\section{Visual Representations of the Computation \& Simulation of Quantum Teleportation}\label{sec:appB}

In this section, we provide examples of descriptive, symbolic and graphical representations of the quantum teleportation algorithm or parts of the algorithm that can be used to target the \textit{Computation \& Simulation} area of expertise.

\subsection{Descriptive} Considering the \textit{Computation \& Simulation} area of expertise, one can describe the algorithm in terms of the quantum gates that are used, which is also done, e.g. in \cite{Nielsen2011-in}. The first and second qubit are initialised in the state $\ket{0}$, while the third qubit is in an arbitrary state to be teleported. Then, the first and second qubit are entangled via a Hadamard gate on the first qubit and a CNOT gate with the first as control and the second as target. The same is done to entangle the second with the third qubit. Then, the first and second qubit are measured to a classical register. In order for the first qubit to be in the original state of the third qubit, quantum gates have to be applied depending on the measurement result, using classical if-operations. If the measurement result of the second qubit is 1, a $Z$-gate has to be applied. If the measurement result of the first qubit is 1, an $X$-gate has to be applied. The order matters only by a global phase.

\subsection{Symbolic} As symbolic representation, the same as in \ref{sec:appA} can be chosen. 

\subsection{Graphical} 

\begin{figure}[htb]
    \centering
    \includegraphics[width=0.9\columnwidth]{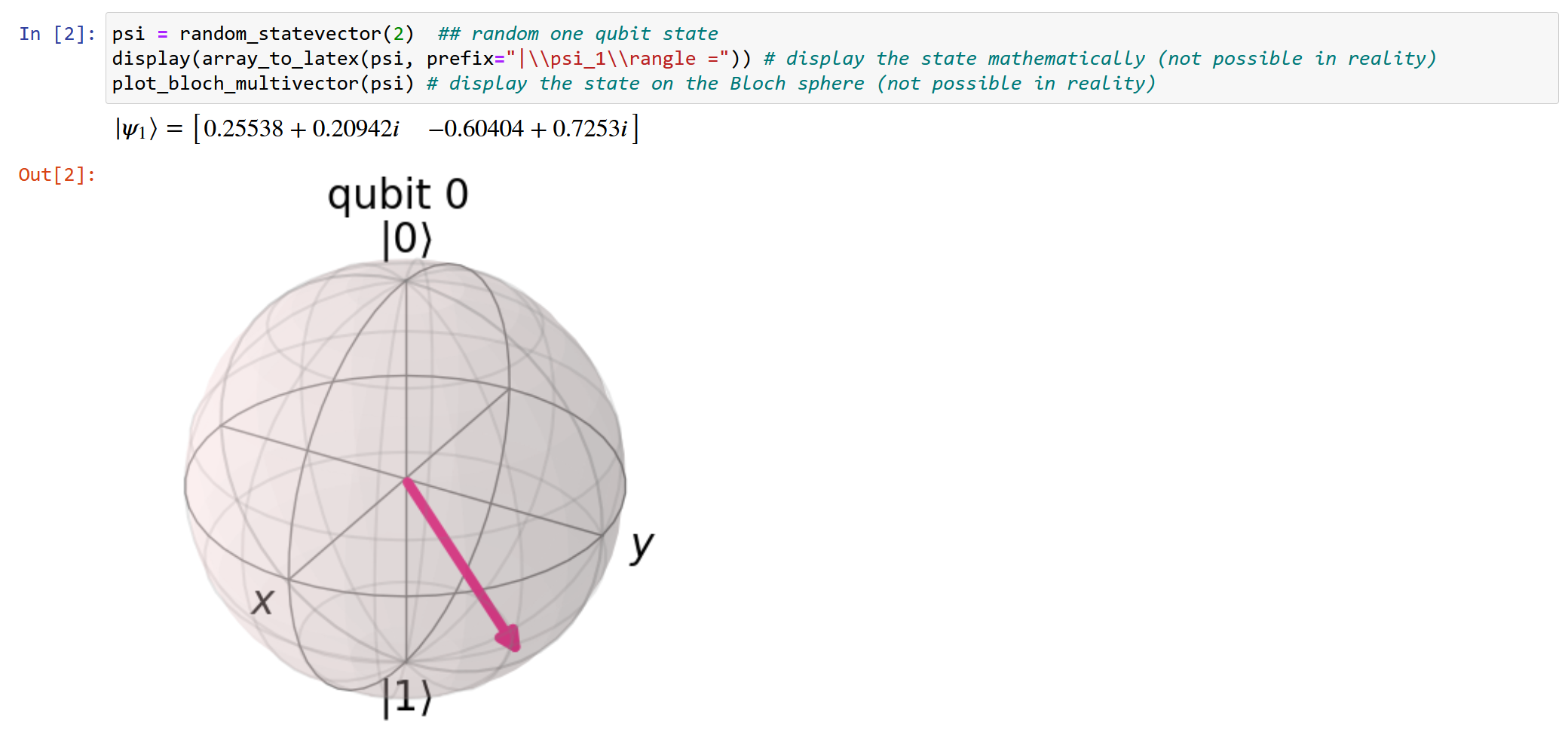}
    \includegraphics[width=0.85\columnwidth]{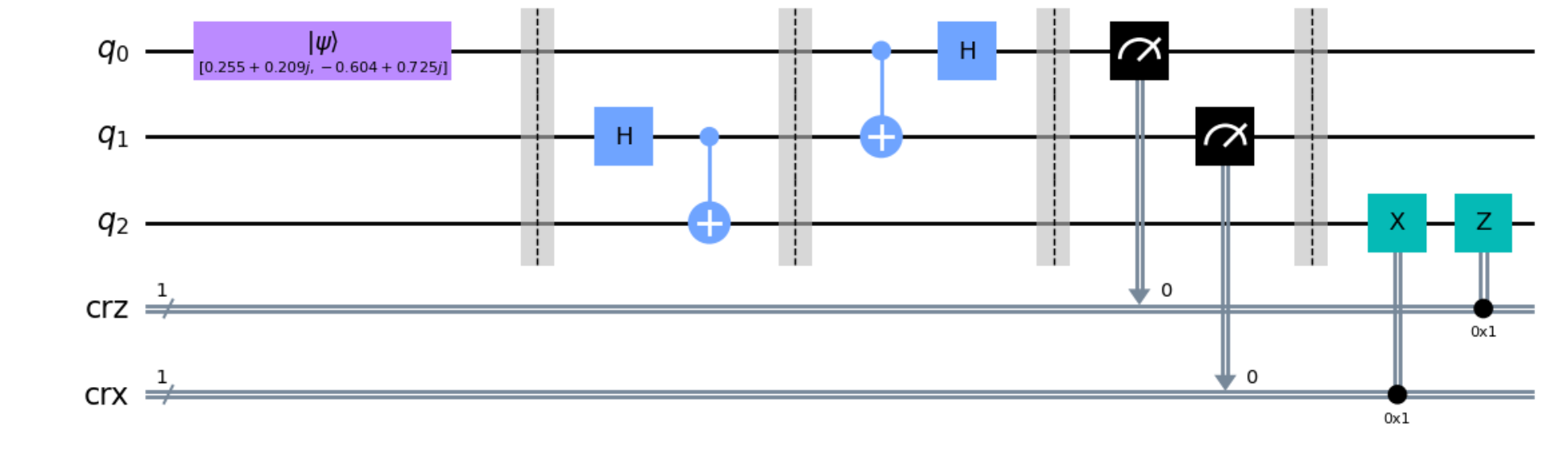}
    \includegraphics[width=0.9\columnwidth]{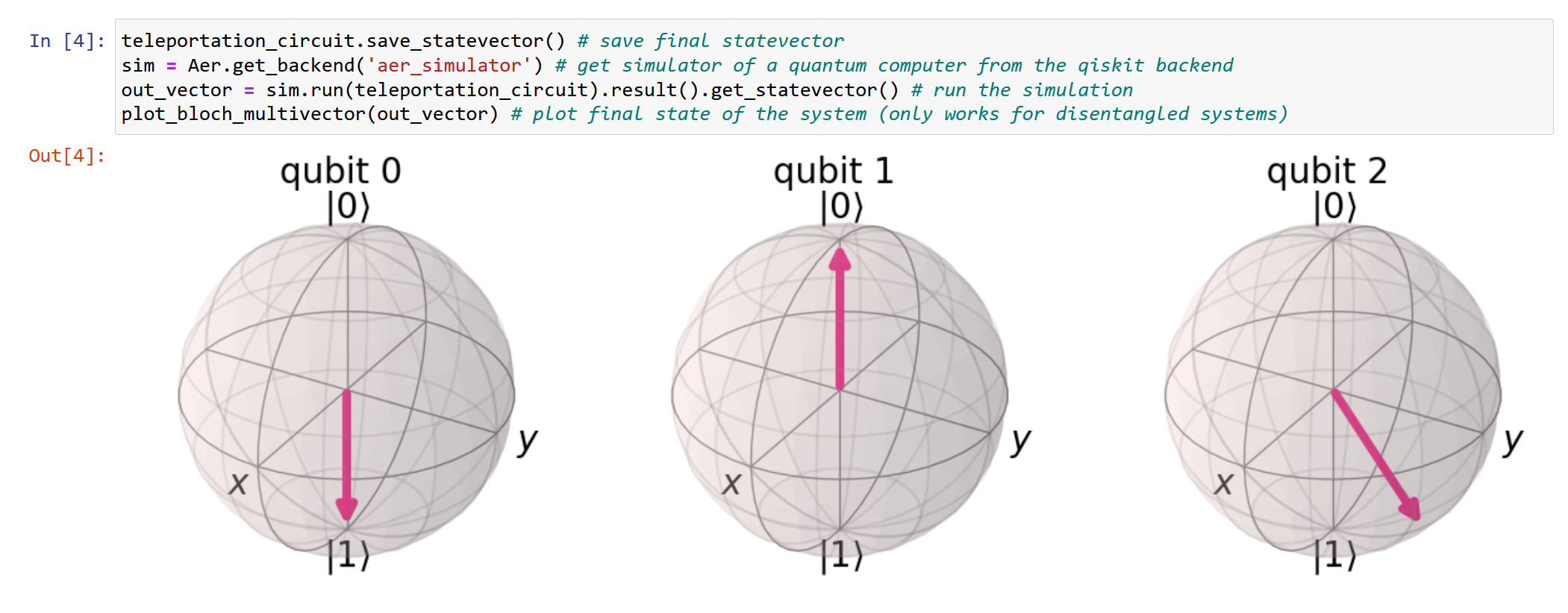}
    \caption{Excerpts from a Python notebook for quantum teleportation in Qiskit. It utilises the circuit representation and the Bloch sphere for visualisation of the computer code and the underlying processes. The first Bloch sphere can be compared to the Bloch sphere of qubit 2 below the code to show that the computer code has functioned properly. Note that this depiction of a quantum state is not possible in reality. Many measurements have to show that a quantum teleportation has worked successfully. Comments are used to make the code more readable. The whole notebook can be found in the supplementary files. Comments in Python code as shown in Fig. 10 are also a form of a descriptive representation of the underlying processes that supports the rather symbolic representation of Python functions.
    }
    \label{fig:QT_comp}
\end{figure}

Computer code is also a form of using \acrlong{mer}. Python functions can be called symbolic, while comments and pictures are supportive representations useful for understanding the code and bug fixing. Some excerpts from a Python notebook for quantum teleportation are depicted in Fig. \ref{fig:QT_comp}.

\section{Visual Representations of Experimental Implementation \& Real World Application of Quantum Teleportation}\label{sec:appC}

In this section, we provide examples of descriptive, symbolic and graphical representations of the quantum teleportation algorithm or parts of the algorithm that can be used to target the \textit{Experiment \& Real World} area of expertise.

\subsection{Descriptive} One possible experimental approach to quantum teleportation is described in \cite{Bouwmeester1997-qo} in the context of quantum optics, i.e. photons. Initiated by a light pulse in the ultraviolet (UV) range, a pair of entangled photons is created via a process called spontaneous parametric down conversion (SPDC) \cite{PhysRevA.54.5349} in a non-linear crystal in such a way that the polarisations are anti-correlated. When the pair now passes through a polarising beam splitter (PBS), they are spatially separated. One can be sent to Bob (photon 3), the other to Alice (photon 2). The initial UV pulse is sent back through the crystal to create a second pair of entangled photons, one of which will be the photon 1 to be teleported for the sake of the experiment, the other of which will serve as an idler photon, indicating that the others are on the way. Alice superposes photon 1 and photon 2 on another beam splitter. If they constructively interfere at that beam splitter, the result will be that each of two detectors will trigger, showing that they were in an anti-correlated state (due to a phenomenon called Hong-Ou-Mandel interference \cite{Walborn2003-ii}). As Photon 1 and 2 were anti-correlated as well as photon 2 and 3, we know (apart from a phase) that photon 3 is now in the original state of photon 1. This phase is detected by Bob using another PBS.

\subsection{Symbolic} The same symbolic representation as in \ref{sec:appA} can be chosen, but $\ket{0}$ and $\ket{1}$ can be represented as $\ket{\leftrightarrow}$, $\ket{\updownarrow}$ to represent the experimental implementation of qubits as done in, e.g. \cite{Bouwmeester1997-qo}.

\subsection{Graphical}

\begin{figure}[htb]
    \centering
    \includegraphics[width=0.9\columnwidth]{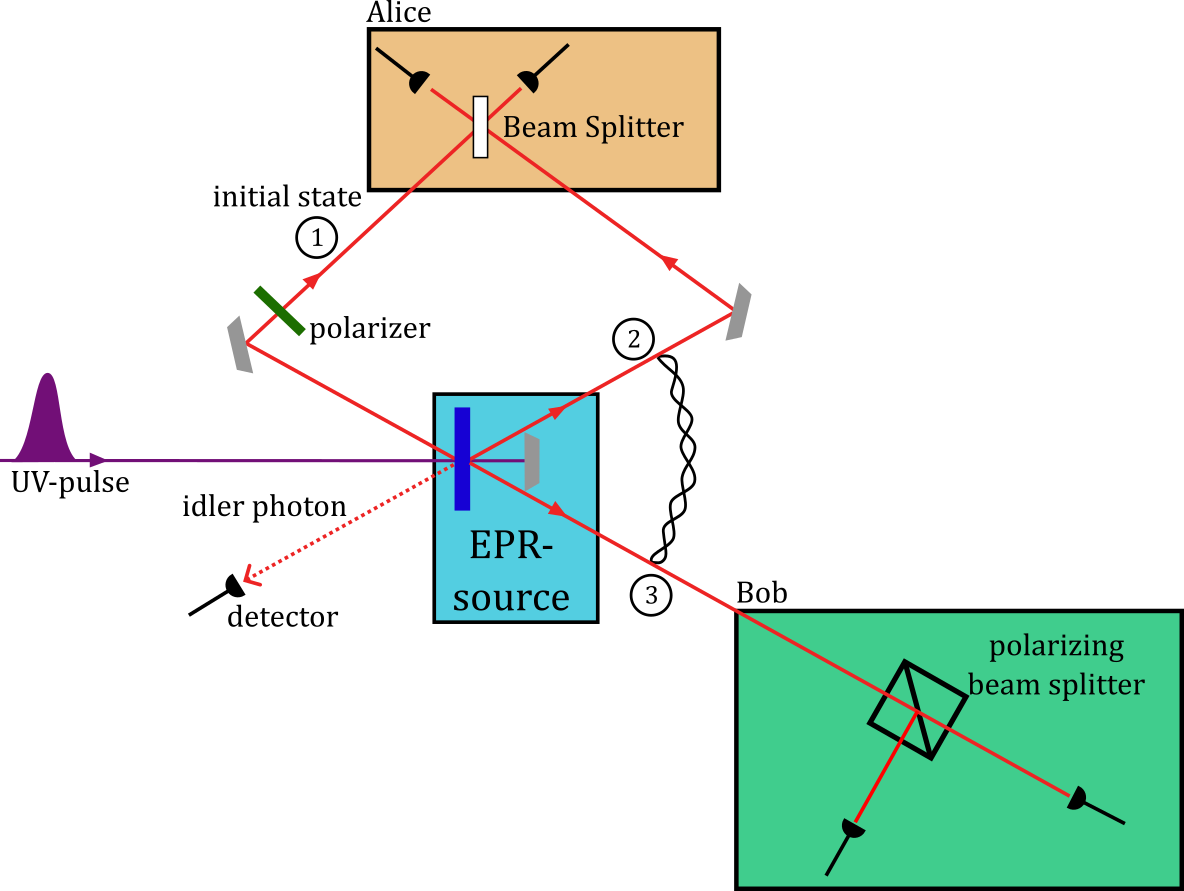}
    \caption{Illustration of experimental quantum teleportation as introduced in \cite{Bouwmeester1997-qo}. An ultraviolet pulse is shot at a nonlinear crystal (blue) and back through the crystal to create four photons labelled 1, 2 and 3 and an idler photon that is used to indicate that the photons were created. Photon 2 and 3 are entangled. Photon 1’s state is manipulated using a polarisor and Alice interferes photon 1 and 2 at a beam splitter. Upon detection of photon 1 and 2 at different detectors, Photon 3 has, apart from a phase, the same polarisation as photon 1 initially had which Bob can confirm using detectors and a polarising beam splitter.
    }
    \label{fig:QT_exp}
\end{figure}

Fig. \ref{fig:QT_exp} shows the explicit experimental setup of quantum teleportation and refers to some process and components that can be used for optical quantum teleportation, namely a nonlinear crystal for SPDC, mirrors, polarisors, (polarising) beam splitters and detectors.

\printglossary[type=\acronymtype]

\printglossary

\section*{References}
\bibliography{main}
\bibliographystyle{unsrt}

\end{document}

%% file: glossary.tex
\makeglossaries

\newglossaryentry{inquiry} 
{
    name=inquiry-based learning, 
    description={A class of pedagogical approaches in which learning is driven by asking and attempting to answer, questions, in a cyclical manner of inquiry} 
}

\newglossaryentry{modality} 
{
    name=modality principle, 
    description={Presenting both visual and audio information to learners can reduce cognitive load as opposed to using just one of the two types \cite{Mayer2005-ed}} 
}

\newglossaryentry{intrinsic} 
{
    name=intrinsic cognitive load, 
    description={Cognitive Load imposed on the learner due to the nature of the task itself \cite{Sweller2011}} 
}

\newglossaryentry{scaffolding} 
{
    name=scaffolding, 
    description={Interactive, temporary support given to learners for tasks that they otherwise would not be able to complete \cite{vandePol2010,Belland2014}} 
}

\newglossaryentry{interdependence} 
{
    name=interdependence, 
    description={in a collaborative work environment, participants are influenced by each other. This interdependence should be positive, i.e. success of one person benefits the others. Positive interdependence can be further broken down into different types of interdependence: goal, reward, resource, role, identity, environmental, fantasy, task and outside enemy \cite{LAAL20131433}} 
}

\newglossaryentry{constructivism} 
{
    name=constructivism, 
    description={The theory that knowledge is constructed by interaction with one's physical and social environment} 
}

\newglossaryentry{Bloom's Taxonomy} 
{
    name=Bloom's Taxonomy, 
    description={A hierarchical model of learning outcomes for science education, ranked from recall, to understanding, evaluation, and creation.} 
}

\newglossaryentry{backwards} 
{
    name=backwards design, 
    description={A backwards design process is a process where one starts with the learning goals and then deduces from there how these learning goals can be achieved \cite{Wiggins2005-oa}} 
}

\newglossaryentry{extraneous} 
{
    name=extraneous cognitive load, 
    description={The cognitive load imposed on the learner due to the design of the task rather than the task itself (which would be \gls{intrinsic}), usually undesirable \cite{Sweller2011}} 
}
\newglossaryentry{aptitude} 
{
    name=aptitude treatment interaction, 
    description={A dependency (interaction) between the learner characteristics (aptitude) and the instruction (treatment) \cite{https://doi.org/10.1002/tea.3660210804}} 
}

\newglossaryentry{redundancy} 
{
    name=redundancy principle, 
    description={Redundant material can hinder learning due to increased \gls{extraneous} \cite{Kalyuga2014-ei}} 
}

\newglossaryentry{expertise} 
{
    name=expertise-reversal effect, 
    description={A form of an \gls{aptitude}, where experts benefit less from certain treatments that are beneficial for beginners, e.g. due to the \gls{redundancy} \cite{doi:10.1207/S15326985EP3801}} 
}

\newglossaryentry{element} 
{
    name=element-interactivity effect, 
    description={For tasks with high \gls{intrinsic} (high element-interactivity), a high \gls{extraneous} can be especially harmful for learning and task solving \cite{doi:10.1207/S15326985EP3801}} 
}

\newglossaryentry{splitattention}{
    name=split attention effect,
    description={\acrshort{mer}s should be integrated to one single source of information rather than split either spatially or temporally to reduce \gls{extraneous} \cite{chandler1992}}
}

\newacronym{cf}{CF}{Competence Framework}

\newacronym{qist}{QIST}{Quantum Information Science \& Technology}

\newacronym{ecl}{ECL}{Extraneous Cognitive Load}

\newacronym{icl}{ICL}{Intrinsic Cognitive Load}

\newacronym{deft}{DeFT}{Designs, Functions, Tasks}

\newacronym{mer}{MER}{Multiple External Representations}

\newacronym{clt}{CLT}{Cognitive Load Theory}

\newacronym{ibl}{IBL}{Inquiry-Based Learning}

\newacronym{qt}{QT}{Quantum Technologies}

\newacronym{stem}{STEM}{Science, Technology, Engineering \& Mathematics}

\newacronym{qctf}{QCTF}{Quantum Curriculum Transformation Framework}